%% file: mnoliver09_v8.tex
\documentclass[useAMS,usenatbib]{mn2e}

\usepackage{graphicx}
\usepackage{color}

\def\spose#1{\hbox to 0pt{#1\hss}} \def\simlt{\mathrel{\spose{\lower
3pt\hbox{$\mathchar"218$}} \raise 2.0pt\hbox{$\mathchar"13C$}}}
\def\simgt{\mathrel{\spose{\lower 3pt\hbox{$\mathchar"218$}} \raise
2.0pt\hbox{$\mathchar"13E$}}}

\def\arcsec{$^{\prime\prime}$}

\def\apj{{ApJ.}}
\def\apjs{{ApJS.}}
\def\apjl{{ApJL.}}
\def\nat{{Nature.}}
\def\mnras{{MNRAS}}

\def\aap{{A\&A}}
\def\aj{{AJ}}

\title[Specific star-formation in the FIR]{Specific star-formation and the relation to stellar mass from $0<z<2$ as seen in the far-infrared at 70 and 160\micron}
\author[S.J. Oliver et al.]{Seb Oliver$^{1}$\thanks{S.Oliver@Sussex.ac.uk}, 
M. Frost$^1$,
D. Farrah$^1$, 
E. Gonzalez-Solares$^2$,
D.L. Shupe$^3$,
\newauthor 
B. Henriques$^1$,
I. Roseboom$^1$,
A. Afonso Luis$^4$,
T.S.R. Babbedge$^5$,
D. Frayer$^6$,
\newauthor 
C. Lencz$^1$,
C.J. Lonsdale$^{7}$,
F. Masci$^{3}$
D. Padgett$^{3}$,
M. Polletta$^8$,
\newauthor 
M. Rowan-Robinson$^5$, 
B. Siana$^6$,
H.E. Smith$^9$, 
J. A. Surace$^6$, 
M. Vaccari$^{10}$
 \\
$^{1}$Astronomy Centre, Dept. of Physics \& Astronomy, University of Sussex, Brighton, BN1 9QH, UK\\
$^2$Institute of Astronomy, University of Cambridge, Madingley Road, Cambridge, CB3 0HA, UK.\\
$^3$Infrared Processing and Analysis Center, MS 100-22, California Institute of Technology, JPL, Pasadena, CA 91125, USA\\
$^4$Institute de Astrofisica de Canarias, C/ Via Lactea s/n, E-38200 La Laguna, Spain,\\
$^5$Astrophysics Group, Blackett Laboratory, Imperial College London, Prince Consort Road, London SW7 2BW, UK.\\
$^6$Spitzer Science Center, MS 220--6,
California Institute of Technology, Jet Propulsion Laboratory, Pasadena, CA 91125, USA.\\
$^7$University of Virginia, 530 McCormick Road, Charlottesville, VA 22094, USA \\
$^8$IASF-INAF Milano, via E. Bassini 15, Milano 20133, Italy.\\
$^9$Center for Astrophysics \& Space Sciences, University of California San Diego, La Jolla, CA 92093-0424, USA.\\
$^{10}$Dipartimento di Astronomia, Universita di Padova, vic. Osservatorio, 3, 35122 Padova, Italy \\
}

\begin{document}

\date{Accepted 200*  **. Received 2008 October *; in original form 2008 September X}

\pagerange{\pageref{firstpage}--\pageref{lastpage}} \pubyear{2006}

\maketitle

\label{firstpage}

\begin{abstract}
We use the Spitzer Wide-area InfraRed Extragalactic Legacy Survey (SWIRE) to explore the  specific star-formation activity of galaxies and their evolution near the peak of the cosmic far-infrared background at 70 and 160\micron.  We use a stacking analysis to determine the mean far-infrared properties of
well defined subsets of galaxies at flux levels well below the far-infrared catalogue detection limits of SWIRE and
other Spitzer surveys.  We tabulate the contribution of different subsets of galaxies to the far-infrared background at 70$\micron\ $ and 160$\micron$.  These long wavelengths provide a good constraint on the bolometric obscured emission. The large area provides good constraints at low $z$ and in finer redshift bins than previous work. At all redshifts we find that the specific far-infrared  luminosity decreases with increasing mass, following a trend $L_{\rm FIR}/M_* \propto M_* ^\beta$ with $\beta =-0.38\pm0.14$. This is a more continuous change than expected from the  \cite{Delucia2007} semi-analytic model suggesting modifications to the feedback prescriptions.  We see an increase in the specific far-infrared  luminosity by about a factor of $\sim100$ from $0<z<2$ and find that the specific far infrared luminosity evolves as
$(1+z)^{\alpha}$ with $\alpha=4.4\pm 0.3$ for galaxies with $10.5<\log_{10} M_*/M_\odot\le12$.  This is considerably steeper than the \cite{Delucia2007} semi-analytic model ($\alpha\sim2.5$).  When separating galaxies into early and late types on the basis of the optical/IR spectral energy distributions we find that the decrease in specific far-infrared  luminosity with stellar mass is stronger in early type galaxies ($\beta\sim-0.46$), while late type galaxies exhibit a flatter trend ($\beta\sim-0.15$).  The evolution is strong for both classes but stronger for the early type galaxies.  The early types show a trend of decreasing strength of evolution as we move from lower to higher masses while the evolution of the late type galaxies has little dependence on stellar mass.  We suggest that in late-type galaxies we are seeing a consistently declining specific star-formation rate $\alpha=3.36\pm0.16$ through a common phenomenon e.g. exhaustion of gas supply i.e. not systematically dependent on the local properties of the galaxy. 

\end{abstract}

\begin{keywords}
galaxies: evolution -- infrared: galaxies -- surveys
\end{keywords}

\section{Introduction}
A fundamental goal of modern astronomy is to understand the processes driving the formation and evolution of galaxies. A key issue is the relationship between the assembly of galaxies, and the formation history of the stars within those galaxies. A galaxy can increase its stellar mass through the accrual of stars in a `dry' merger where the merger does not trigger new star formation or directly through star formation triggered by a merger or some other process.  The star formation rates are determined by a variety of factors including the triggering mechanisms, the supply of gas and the feedback processes. The contribution from all of these three processes to stellar mass buildup in galaxies is subtle, and has been studied in numerous optical/near-IR photometric and spectroscopic surveys. These surveys have demonstrated that, as we increase in redshift, there is a strong dependency on at least two parameters; galaxy mass, and local environment.

Galaxy mass is thought to play a important role, at least at $z\simlt1$ \citep{cas07}. At low redshifts ($z\simlt0.2$), ongoing star formation in massive galaxies is almost entirely absent. Extreme levels of star formation are found rarely and in many cases are triggered by interactions and mergers.  Moderate star formation is probably triggered by internal processes \citep{owe07,melbourne08}. As we move to higher redshifts however, this picture changes. Overall, the galaxy stellar mass function at high masses evolves fairly slowly up to $z\sim0.9$, and then more rapidly up to at least $z\sim2.5$, suggesting that the majority of stellar mass assembly took place at $z\simgt1$ \citep{poz07,feu07}. Massive galaxies show little evidence for stellar mass assembly via either star formation or dry mergers at $z\simlt0.7$, while lower mass systems harbour ongoing star formation at all redshifts, lending support to the idea of `downsizing', in which more massive galaxies form most of their stars at high redshifts. There is also some evidence for `dry' mergers \citep{bel06}. Finally there is evidence that higher mass galaxies have lower {\it specific} star formation rates (sSFR, i.e the star formation rate per unit stellar mass) than do lower mass systems over a very wide redshift range, possibly up to $z\sim4$, suggesting that lower mass galaxies  form stars more efficiently \citep{bau05,feu05,zhe07}, although the sSFRs of massive galaxies appear to increase rapidly with increasing redshift. 

Local environment also has a significant effect. In the local Universe we see a distinct environmental segregation, in which galaxies in rich environments show much lower star formation rates than do galaxies in the field \citep{cas07,zau07,coo07}. At higher redshifts however this trend is reversed; at $z\sim1$ the average star formation rate increases with increasing environmental density, as does specific star formation rate \citep{elb07}, and morphological evolution appears to be more rapid in dense environments than in the field \citep{cap07}. Star forming galaxies at $z\sim1$ are in richer regions than seen in the local Universe \citep{far06, maglio07}, and there is evidence that star formation gradually shifts to lower density regions from $z\sim1.5$ to z=0 \citep{dlt07}, thus providing a natural explanation for the local environmental segregation.  Since environmental effects are so important we can only understand the universal properties when averaging over a representative sample of environments.

An important consideration is the feedback from supernovae or AGN which is required to suppress star formation in semi-analytic models (SAMs).  A growing consensus is that the models require AGN feedback to suppress the star formation
in massive objects \citep{croton,bower} and this is in fact the dominant mechanism in \citet{Delucia2007}. 

The specific star-formation rate is a particularly useful probe of these processes.  It measures the ratio between the current star-formation rate and the historically averaged rate (e.g. \citealt{Brinchmann,Walcher}).  Enhanced star-formation rates arising simply from  an increase in gas reservoirs achieved through  `dry mergers' will not affect the sSFR, which is mainly sensitive to the triggering, fuelling and feedback i.e. the overall star-formation efficiency. Thus the sSFR provides some de-coupling of the merging and other phenomena.

There remains however a significant problem with these studies. The discovery of a strong cosmic infrared background (CIRB) by COBE \citep{pug96,hauser98}, and its subsequent (partial) resolution into a huge population of obscured star-forming galaxies at $z\simgt1$ by ISO \citep{rr97,dol,mann02,ver05}, Spitzer \citep{lef05}, and in the sub-mm \citep{hug,eal,scot,bor,mor}, demonstrated clearly that a large fraction of the total star formation at high redshifts is heavily shrouded in dust, and therefore impossible to detect at optical or near-IR wavelengths. This applies to even moderately luminous systems at $z\sim1$ (e.g. \citealt{lef05}, their figure 14). Therefore, optical/near-IR surveys to probe stellar mass assembly at high redshift miss a significant fraction of ongoing star formation.

The Spitzer space telescope \citep{wer04} has allowed us to make great progress in understanding these obscured systems. However, even the recent surveys that combine optical data with 24$\mu$m data do not solve this problem; the peak rest-frame emission from most IR-luminous galaxies is in the range 40-100$\mu$m, but at $1<z<2$ the 24$\mu$m Spitzer band probes rest frame 8-12$\mu$m, a region which can be highly contaminated with a range of features (i.e. PAH emission, Si absorption etc).
Spitzer does however have two longer wavelength channels, at 70 and 160$\mu$m. The resolution and sensitivity of these channels are insufficient to directly resolve the far-infrared (FIR) background into the individual galaxies that produce it (e.g. \citealt{dol04}), but techniques exist to alleviate this. One of these techniques is known as `stacking'. 

To understand galaxy evolution it is thus essential to understand the obscured specific star-formation rate of galaxies (and massive galaxies in particular) over a representative range of galaxy environments.
In this paper, we use a stacking technique to measure the specific far infrared luminosity at 70 and 160\micron\ as a function of galaxy mass and redshift. To probe the highest stellar mass objects, which are rare, we use a large survey area.  Although we don't explore the variation with environment in this paper this large survey area also means we can be confident that we have covered a representative range of environments. 

In Section 2 we discuss the samples. In Section 3 we describe our stacking technique. In Section 4 we present the results, which we discuss in Section 5, before concluding. We assume a spatially flat cosmology with $H_{0}=100\,h\,{\rm km}\; {\rm s}^{-1}$ Mpc$^{-1}$, $h=0.7$, $\Omega=1$, and $\Omega_{m}=0.3$  and all magnitudes are Vega magnitudes.

\section{Catalogues and Maps}
\subsection{SWIRE MIPS Maps}\label{sec:mipsmaps}

The Spitzer Extragalactic Legacy Survey (SWIRE, \citealt{lonsdale03}, \citealt{lonsdale04})  observed 49 square degrees in six fields 
(CDFS, Elais-N1, Elais-N2, Elais-S1, Lockman and XMM-LSS) using the seven primary Spitzer imaging bands (3.6 to 160\micron). The SWIRE 70\micron\ and 160\micron\ maps used here were observed using MIPS medium scans as described in \cite{Shupe}.  The
70\micron\  maps were made from Basic Calibrated Data (BCD) images produced  by the
SSC pipelines, after applying time filtering and column filtering  following the
prescription in \citet{Frayer06}, and are part of our Data Release 4.  The filtered BCDs were  mosaicked
using the MOPEX package \citep{mopex}.  The 160\micron\  maps  were made by
mosaicking the filtered BCD images produced by the SSC pipelines and  correcting
for a systematic 5\arcsec\ pointing offset.  The maps are calibrated  in units of
surface brightness (MJy/sr). To bring the calibration in-line with the facility calibration appropriate for the ``S13" pipeline processing (May 2006) we
scale the maps up by a factor of 1.107 at 70\micron.  The 160\micron\ maps for CDFS, XMM-LSS and Elais-N2 were multiplied by 1.064 while those of Lockman and Elais-N1 were unchanged (having already been reprocessed with the latest calibration).  I.e. our map calibration is 702 MJy/sr per MIPS-70 unit and 44.7 MJy/sr per MIPS-160 unit \citep{Gordon2006, Stansberry2006}.

We estimate point-source fluxes, $f$ (in mJy), by fitting the point source response function (PRF) to the map intensity $I$ 
(in MJy sr$^{-1}$).   Our PRF is
based on the default MIPS PRFs as recommended for use with Apex from the Spitzer Science Centre\footnote{http://ssc.spitzer.caltech.edu/mips/dh/}.  These PRFs have the same pixel size as the SWIRE maps i.e. 4\arcsec at 70\micron\ and 8\arcsec at 160\micron\ and are over-sampled by a factor of 4 i.e. 1\arcsec(for 70\micron) and 2\arcsec (for 160\micron). These PRFs are not identical to the ones used for the SWIRE catalogue extraction but the resulting difference calibration is less than 5\% and smaller than the absolute calibration uncertainty.

We define an effective beam size $\Omega_{\rm beam}=f/I_0=\int P\,d\Omega/P_0$ (where $P$ is the PRF, $d\Omega$ a solid angle and the zero subscript indicates the peak). We obtain $\Omega_{\rm beam}=13.1$~nSr at 70$\mu$m and $\Omega_{\rm beam}=61.5$~nSr at 160$\mu$m. 

Finally, for consistency with the SWIRE catalogues we apply a colour correction to convert from the standard calibration for a 10,000 K blackbody to constant $\nu f_\nu$ more appropriate for galaxies.   We therefore multiply the 70 and 160\micron\ fluxes by 1.09 and 1.043 respectively.

We have checked our calibration by comparing the map intensity at the position of catalogued MIPS sources \citep{AfonsoLuis06}  and find a good agreement (within the absolute calibration uncertainties). 

We note that the absolute SSC calibration is good to an accuracy of  7\% for 70\micron\ and 12\% for 160\micron\ \citep{Gordon2006, Stansberry2006}.

\subsection{SWIRE optical/IR band-merged catalogues}
The fields Elais-N1, Elais-N2, Lockman and CDFS currently have the most homogenous and well understood SWIRE data. The
 catalogues we are using are those that were released as part of SWIRE Data Release 4 \citep{2004yCat.2255....0S}.  
All IRAC and MIPS catalogues have been ``bandmerged" i.e. independent catalogues from the seven different bands (3.6, 4.5, 5.8, 8.0, 24, 70 and 160\micron) have been cross-matched to produce one master catalogue.
In Elais-N1 and Elais-N2 we have 5-band ($UÕgÕrÕiÕZÕ$) photometry from the Wide Field Survey (WFS, \citealt{mcmahon}, 
\citealt{irwinlewis01}).
 In the
Lockman Hole we have 3-band photometry ($gÕrÕiÕ$) from the SWIRE photometry programme,
with some additional U-band photometry.  In Chandra Deep Field South (CDFS)
we have 3-band ($gÕrÕiÕ$) photometry, also from the SWIRE
photometry programme.  Good optical data exists in other SWIRE fields, notably the CFHTLS\footnote{http://www.cfht.hawaii.edu/Science/CFHTLS/} data in XMM-LSS
and data from ESO WFI surveys in Elais-S1 (\citealt{berta06}, \citealt{berta08}), however, these fields have not been used for much of our analysis as it is known that our photometric redshift estimates are not as good in these fields.
 
\citet{frost09} have estimated completeness limits of
 $u=22.7, g=23.8, r=23.2, i=22.4, z=21.2$ Vega magnitudes and  $f_{3.6\micron}=10, f_{4.5\micron}=15$mJy and we use those estimates where required in our following analysis.

\subsection{Photometric redshifts}
\label{subsec:photo-z}
We use the photometric redshifts given in \cite{RowanRobinson07}.
These use the code I{\scriptsize MP}Z (\citealt{Babbedge04}, \citealt{RowanRobinson03} with updates as described in \citealt{RowanRobinson05} and \citealt{Babbedge06}) which has been extensively tested and applied to SWIRE data.

I{\scriptsize MP}Z  is a template-fitting code that utilises the optical and IRAC 3.6\micron, 4.5\micron\  detections to produce reliable photometric redshifts for both galaxies and AGN, accounting for extinction.   The code has been tested against numerous spectroscopic datasets within the SWIRE fields with optical data down to $r<24$ and spectroscopic redshifts $z\simlt3$ (though with most $z<1.5$).  For all the samples, the mean systematic offset between the photometric and spectroscopic redshifts was found to be negligible to the precision of the photometric redshifts.  For example, the Elais-N1 sample used by \citet{Babbedge06} has a systematic offset of only $\langle{\Delta} z/(1+z)\rangle=+0.0037 $.  
 \cite{RowanRobinson07} provide a detailed analysis of the catastrophic failure rate $\eta$ and photometric accuracy $\sigma_{\rm phot}^2=\langle(\Delta z/(1+z))^2\rangle$ as a function of number of photometric points, $n_{\rm band}$,  magnitude cuts and $\chi^2$ from the fit. 
From their Figure 11 we estimate that we would find $\sigma_{\rm phot}\le5$\% and $\eta \le7$\% with $\chi^2< 5.$ and $n_{\rm band} \ge5$ and $r<24$. We use these constraints for the analysis presented in Section \ref{sec:cirb}, though with a much brighter $r<23.2$ limit.  However, as shown in \citet{frost09} and discussed in Section \ref{sec:selectioneffects} these constraints are difficult to model, so we modify them slightly for the analysis in Section \ref{sec:ssfr}.  The $\chi^2$ constraint does not affect many galaxies and is difficult to adapt so we maintain this. We create a modified $n^\prime_{\rm band}$ measure being the number of bands that exceed the specific completeness thresholds $u=22.7, g=23.8, r=23.2, i=22.4, z=21.2$ and  $f_{3.6\micron}=10,f_{4.5\micron}=15$mJy (as opposed to the more general requirement of a detection).  We adopted a selection of $n^\prime_{\rm band}\ge 4 $ which provided a similar set of galaxies to $n_{\rm band} \ge5$ and thus, presumably, a similar redshift accuracy.

\subsection{Optical classes and stellar mass estimation}
\label{subsec:lum}

We also used the spectral energy distribution (SED) classifications and stellar masses given in  (RR08:  \citealt{RowanRobinson07}).  They adopted a two-pass approach in order to fit the 
photometric redshift,  optical/near-infrared SED and FIR SED making maximum use of the near/mid-IR Spitzer data. The optical and IRAC 3.6 \& 4.5$\mu$m bands are fit first with a range of optical-near IR SEDs based on stellar population synthesis models. The main purpose of this first pass is to determine the level (if any) of excess emission in the Spitzer bands from dust, as well as separate AGN and galaxy spectral types for the second pass. The second pass includes a refitting of the optical SEDs to a finer redshift grid, as well as a fit to the far-IR component of the data with a range of mid-to-far IR templates.
In this work we do not make use of their FIR luminosities or classifications but we note that because of this two-pass method the optical 
luminosity estimates and hence stellar masses should not be biased by any residual FIR contamination in the mid-IR Spitzer bands.

The stellar masses use stellar synthesis templates (see Section 3, Figure 1 and Table 2 of RR08).  
Starting with empirical templates from \cite{yt88} for galaxies of type E, Sab, Sbc, Scd, Sdm, and
from \cite{calzetti92} for starbursts, RR08 used spectroscopic data for 5976 galaxies, for many of which 
they have 10-band photometry from the CFH12K-VIRMOS survey \citep{lefevre04}, to improve these empirical
templates.  The latter were then regenerated to higher resolution using simple stellar populations, 
each weighted by a different SFR and extinguished by a different amount of dust, $A_V$.  
This procedure,
 based on the synthesis code of \citet{poggianti2001}, gave the templates a physical validity.  
Minimization was based on the Adaptive Simulated Annealing algorithm.  Details on this algorithm and on the fitting 
technique are given in \cite{berta04}. 

For each galaxy they estimate 
the rest-frame 3.6 $\mu$m luminosity, $\nu L_{\nu}(3.6)$, in units of $L_{\odot}$, and using the stellar synthesis models
estimated the ratio $(M_*/M_{\odot})/(\nu L_{\nu}(3.6)/L_{\odot})$ to be 38.4, 40.8, 27.6, 35.3, 18.7,
26.7, for types E, Sab, Sbc, Scd, Sdm, sb, respectively.  
[Note:  measuring the 3.6 $\mu$m monochromatic luminosity in total solar units, not in units of the sun's
monochromatic 3.6 $\mu$m luminosity.]  Alternative estimates of $M_*$ using the B-band luminosity agree 
with our preferred method to within 10-20$\%$.  Estimates based on 3.6 $\mu$m should be more reliable, since there is a
better sampling of lower mass stars and less susceptibility to recently formed massive stars.
These mass estimates would be strictly valid only for low redshift.  For higher redshifts the mass-to-light estimates will be lower
since for the oldest stellar populations, $M/L$ varies strongly with age (\citealt{bc93}, see their Figure 3).
This can be approximately modeled using the \citet{berta04} synthesis fits described above, with an 
accuracy of 10$\%$, as 
$(M_*/M_{\odot})/(\nu L_{\nu}(3.6)/L_{\odot})(t) = 50/[a+1.17(t/t_0)^{-0.6}]$
where $t_0$ is the present epoch and $a = 0.15, 0.08, 0.61, 0.26, 1.44, 0.70$ for SED types E, Sab, Sbc, Scd, Sdm and sb, 
respectively.

This approach should correctly capture the different evolutionary behaviour of stellar masses in 
star-forming galaxies and of early-type galaxies.

For each galaxy we have the best fit photo-$z$, the template classification and stellar mass. There are 15 numbered optical classifications 
Ellipticals (1-2), Sab (3), Sbc(5), Scd(7), Sdm (9), star-burst (11), AGN (13-15)   
with types 4, 6, 8, 10 intermediate between the other optical templates. For much of our analysis we exclude the galaxies best fit by AGN templates as their photo-z are poor and the estimates of their star-formation and stellar mass will be strongly contaminated by the AGN.
  
\subsection{Masks}\label{sec:masks}
We have constructed a conservative mask based on those used in the clustering analysis of \cite{frost09}\footnote{http://astronomy.susx.ac.uk/\~\ mif20/masks/index.html}.
Those clustering masks exclude regions where (1) the optical data is insufficient for reliable photo-$z$ determinations (2) the IRAC completeness was low and/or variable and (3) foreground stars were located.  Specifically we include only those optical frames where the $r_{95}$ depth was recorded as 23.2 or better. This $r_{95}$ depth is an estimate of the $r$-band 95\% completeness limit and  $r_{95}=23.2$ is the expected depth for good photometric nights. This means we are selecting good quality optical fields.  Our optical catalogues are also restricted to $r=23.2$. We reject areas where the IRAC coverage was less than 4 pointings, ensuring the completeness is above 90\% and variations are less than 2\% at 10$\mu$Jy.  We also apply a coverage cut of 30 at 24 \micron\  giving completeness $>90$\% and variation of a couple of percent for 24 \micron\ sources above 400 $\mu$Jy.
To mask stars we follow the method of \cite{waddington07} in which 2MASS \citep{skrutskie06} point source catalogue sources with $K<12$ are identified to be stars and a circular mask of radius, R, with $\log_{10}(R/^{\prime\prime}) = 3.1 - 0.16 K$ is applied.  
We additionally exclude regions where the MIPS coverage is poor relative to the majority of the data with a threshold chosen by examining the histogram of the coverage. For the MIPS 70\micron\  we use a coverage threshold of 12, while for the 160\micron\ we use a coverage threshold of 3.  

Although parts of our analysis could be carried out with less conservative masks we choose to apply the same mask to all the work in this paper so we are always comparing sources in the same area of sky.
With our aggressive masking the unmasked areas of the four fields Elais-N1, Elais-N2, Lockman and CDFS
are  $\Omega=4.33, 2.41, 2.75$  and 1.85 sq. deg., totalling 11.33 sq. deg..

\subsection{Sub-sample selection}\label{sec:bin}
\begin{figure*}
\includegraphics[width=16cm]{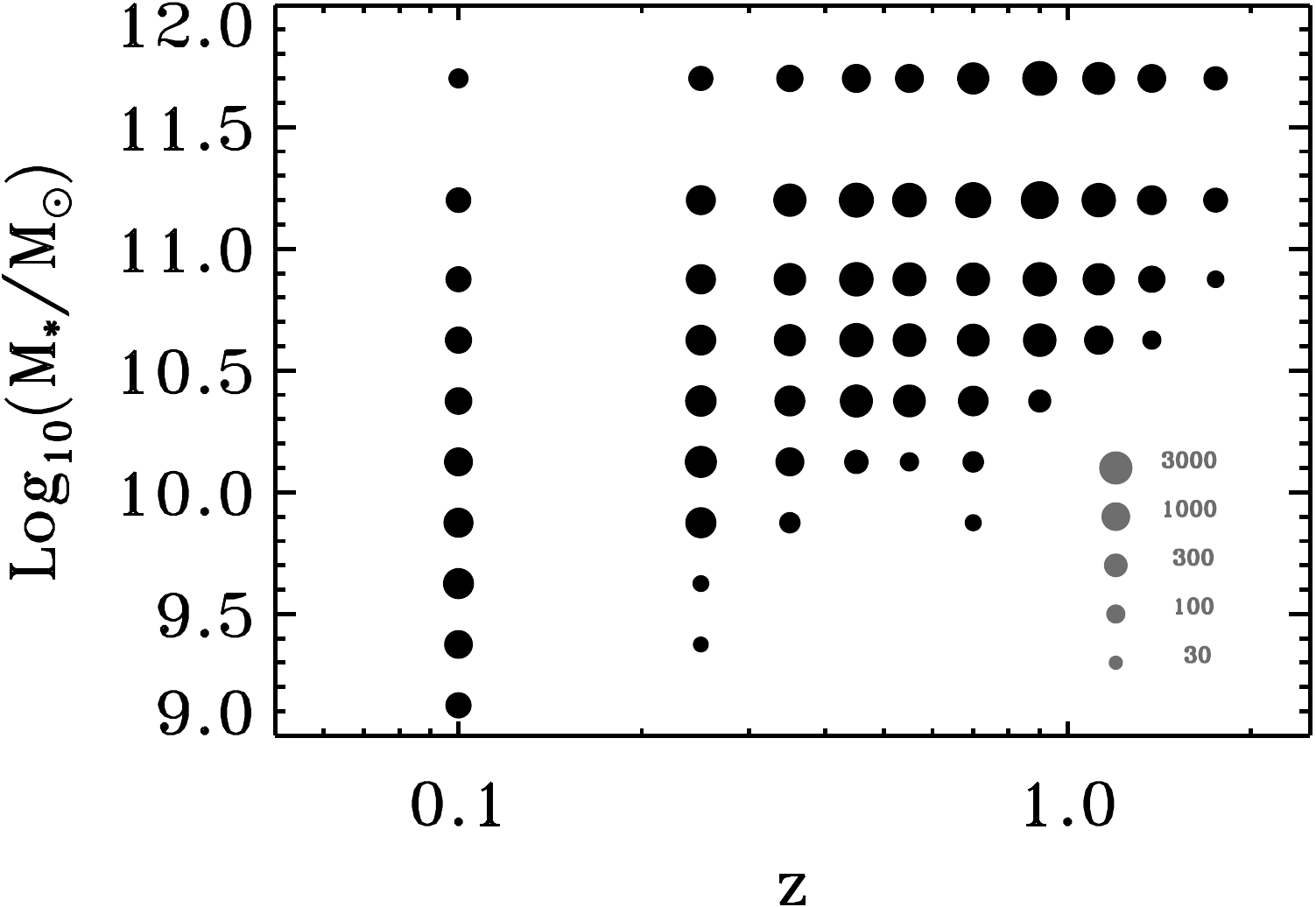}
\caption{Stellar mass vs photo-z plane for groups in our analysis. Each point represents the centre of a $M_*,z$ cell with the size representing the number of objects in the cell according to the code in the legend.}\label{fig:lz}
\end{figure*}

For part of our analysis we divide the sample into stellar mass and redshift ($M_{\rm *},z$) cells over a range $0\le z\le 2.0$ and  $9.0\le \log M_{\rm *}/M_\odot \le 12$.
 Cell sizes were selected to provide relatively uniform number of sources in each cell and thus a reasonable balance between signal-to-noise and resolution on a scale that is easy to compare with other data and models.  Our cell boundaries are
 $z = 0., 0.2, 0.3, 0.4, 0.5, 0.6, 0.8, 1.0, 1.25, 1.5,2$
 and  $\log M_{\rm *}/M_\odot= 9, 9.25, 9.5, 9.75, 10., 10.25, 10.5, 10.75, 11, 11.4, 12$.
 The stellar mass and redshift distribution of the cells is shown in Figure~\ref{fig:lz}.

With redshift slices of $\Delta z\approx0.2$ and area of $\Omega=11.33$ sq. deg. our sample covers a co-moving volume of 6.3$\times 10^{6}h^{-3}$Mpc$^{3}$ at $z=1$.  Numerical simulations \citep{mowhite} predict that this volume is sufficient to include the progenitors of 6 of today's $10^{15}M_\odot$ clusters, which have a co-moving number density of $10^{-6}h^3$Mpc$^{-3}$, arguing that we probe a fair sample of the Universe.

\subsection{Selection effects}\label{sec:selectioneffects}
Our sample and photo-$z$ selection criteria are complex so we need to be cautious about selection effects. In summary, our selections for the stellar mass analysis are: $f_{36\micron}>10\mu{\rm Jy}$, $r<23.2$, $n_{\rm band}^\prime\ge4$, $\chi^2<5$, template classification $j_2<13$ and lying outside the mask defined in Section \ref{sec:masks}. 

We are primarily investigating the mean FIR luminosity to stellar mass ratio, i.e. the specific FIR luminosity, and are not concerned with  the total number of sources or the luminosity density.  Thus 
in any $M_*,z$ cell we only need to worry about selection effects that affect sub-populations within the cell in different ways. In other words our estimate of the mean specific FIR luminosity in a cell would be unaffected if we randomly exclude 50\% of all galaxies in the cell but would be affected if we exclude 50\% of the most FIR luminous and none of the others.  

The $\chi^2$ selection might affect some galaxies types more than others, however, with a broad range of templates, carefully modified to fit the SWIRE populations, we expect this cut will have a smaller effect than the flux related criteria.

The number of bands criteria and the flux cuts do need to be considered. Since different galaxies have different K-corrections this introduces some differential effects. As noted by \citet{frost09}, in a cell of fixed stellar mass (or 3.6 \micron\ luminosity) and redshift the optical cuts will exclude redder galaxies. As these galaxies typically have lower specific star-formation rates ignoring this effect would artificially inflate our specific FIR luminosity estimates.  To account for this we weight the galaxies to account for this incompleteness and flag as incomplete cells where our incompleteness corrections may be inadequate.

For all cells we use a $V_{\rm max}$ weighting. The model templates give us a K-correction in every band. For each galaxy we use our measured K-correction and observed magnitude to calculate the maximum observable redshift $z_{{\rm max},i}$ in each band, $i$.  We then compute $z_{\rm max}$ taking into account our $3.6\micron, r$-band and $n^\prime_{\rm band}$ constraints.  The available volume can then be calculated taking into account the cell limits and $z_{\rm max}$.  We also flag as incomplete any cells in which for any template; (1) we found no examples of galaxies best fit by that template and (2) a galaxy with that template could not, in principle, have been found in the cell, given the quoted selection limits.  Later, we also exclude any cells with fewer than 100 galaxies.


\section{Stacking method}\label{sec:method}
Stacking analysis co-adds the signal in a map at the position of a {\em class} of galaxies, allowing for the reduction of noise associated with measurements of individual galaxies, whether this be confusion noise (e.g. \citealt{con74}) or instrumental/background noise. The level of noise reduction is governed by a number of factors, including the ability to categorise the {\em target} catalogues into groups with similar properties and the resulting number of targets on which the stack is performed. Stacking has been used at sub-mm wavelengths (e.g. \citealt{pea00,dye07,tak07,serj08}), and has successfully been applied to Spitzer data \citep{dol06}. Stacking has also been used to investigate the far-infrared spectral energy distributions of Spitzer galaxies in small fields at $z\sim0.7$ \citep{zhe07b} and  $1.5<z<2.5$ \citep{pap07}.

The conventional stacking technique is to extract a region from the map around each target source and stack these together to produce an average image. A mean background is subtracted; usually
estimated either from the global map or from the extremities of the average image. The average flux is then 
determined by either calculating the total flux in some aperture or  fitting the point-spread 
function to the image.   If the noise in the original maps is uncorrelated with the
target sources (being either instrumental noise, background thermal noise, confusion noise
from unrelated galaxies, or foregrounds such as Galactic cirrus) then the noise in the average map will
be reduced by a factor of $\sqrt{N}$ where $N$ is the number of targets. 
 Furthermore, if $N$ is large then the noise will approach a Gaussian distribution due to the central limit theorem. 
Under the assumption that the targets are isolated and point-like at the telescope resolution  a suitably weighted 
point-source profile fit is the optimal estimator of the average flux of the targets.  

An additional contribution to the average flux will come from sources that are spatially correlated with the 
target galaxies.  This is a difficult contribution to model as it depends on the correlation function of
galaxies which may be luminosity or type dependent. We discuss this further in Section~\ref{sec:sims}.  
In order to minimize this bias  we fit the point-source profile and a constant background simultaneously 
over a limited radius.   (A similar technique for source detection and photometry has been discussed by \citealt{2007ApJ...661.1339S}).
The simultaneous fit means that the background is estimated from a local area that includes the region under the source, in contrast to methods using a sky annulus which exclude the source region.
The 
point-source fitting and limited radius means that the central parts of the image, where the source flux dominates any correlated background, have a greater weight.  

We performed the fit in 1D using a radial point-source profile $P(r)$  estimated from the 2D calibration files. The two parameter intensity profile function $I(r)=I_0\, P(r)/P_0 + b$ was fit, 
limiting the data to that within the first Airy disk minimum. We used a minimum $\chi^2$ fit with errors in the mean image intensity estimated from the scatter between the individual images in the stack.  This would be optimal if the population variation is small and the errors are un-correlated, so it probably under-weights the central pixels but is adequate for our purposes.   From $I_0$ we deduce the flux, $f$, using the effective beam calibration factor, $\Omega_{\rm beam}$, given earlier. Looking at the $\chi^2$ we find reasonable fits whenever we have 9 or more galaxies.  

It is simple to use the central limiting behaviour of the stacking technique to get a good estimate of statistical error in the flux.  However, we are concerned about systematic noise terms and so will use an error calculation based on the variation from field to field (see Section \ref{sec:errors}).  Our technique can easily be extended to include correlated errors, to accommodate a model for any background (e.g. one from correlated sources) and also to provide an estimate for the variation in the population flux as well as the mean.

\subsection{Simulations}\label{sec:sims}
From the beam sizes calculated above we see that the density of beams on the sky is 21000 and 4700 per square degree at 70\micron\ and 160 \micron, respectively.  Comparing these with Table~\ref{tab:cirb} we can see the number of sources per beam can be high, particularly at 160\micron. We should thus be concerned about confusion i.e. where correlated neighbouring sources spuriously increase the stacked flux.  

\begin{figure*}
\includegraphics[width=16cm]{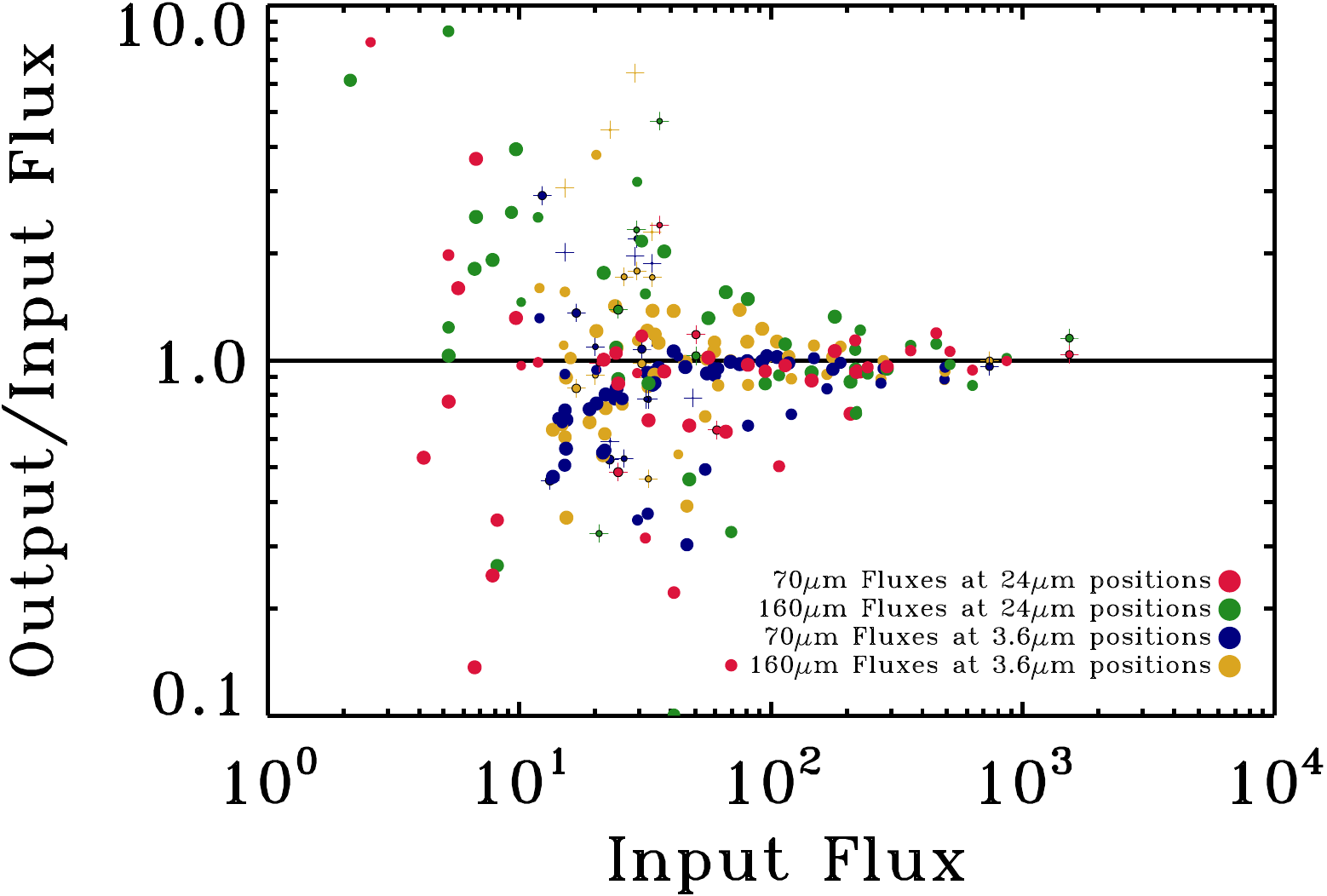}
\caption{Ratio of the output to input flux ratios for simulations of our stacked samples. Flux units are arbitrary. We use the stack samples from the Elais-N1 field (our largest field, the total sample has over twice this area). The colours of the symbols indicate whether we placed the FIR fluxes at the location of 3.6 \micron\ or 24 \micron\ sources and whether we used the 70 or 160 \micron\ beam. Symbol sizes relate to sample sizes in the same way as Figure \protect\ref{fig:lz} except the symbol sizes are reduced by a factor of two.  Cells with fewer than 10 sources are marked with crosses.}\label{fig:sims}
\end{figure*}

Our method aims to mitigate this problem by using the simultaneous source and background fitting.  To test this we have run some simulations. We need to investigate the behaviour in high and low source densities regimes and to include galaxy clustering.  Our approach is to use the real catalogue positions.  For the low source density case we take the sources with $S_{\rm 24}>400\mu$Jy. We 
are not interested in the absolute fluxes, so we model the long-wavelength fluxes as $S_{\rm 70}=S_{\rm 160}=S_{24}$. We
insert these fluxes at the catalogue positions and convolve with the corresponding 2D point response function. For the high source density case we repeat this using sources with $S_{\rm 3.6}>10\mu$Jy.  We
then undertake our stacking analysis using the same stellar mass and redshift cells.  We do not add noise to the simulations as we are concerned here with systematic biases from confusion.  The results are shown in Figure~\ref{fig:sims}.

These simulations show that the uncertainties increase as the mean flux in the stack decreases. The results are roughly similar for the two different sample densities and the two different beam sizes. At any given flux the uncertainties appear to be larger for smaller stack samples.  If we only consider the larger stack samples then there is some indication of systematic {\em underestimation} of the simulated flux at faint fluxes. This effect appears to be less than 0.5 dex.    At first glance this  is surprising since our na\"{i}ve expectation was that the correlated signal would increase the fluxes.  This suggests that our background estimation is biased upwards by the correlated flux. In principle we could model this bias.  However, our simulation is limited as (1) we have assumed a direct correlation between the long wavelength fluxes and short wavelength fluxes and (2) we have ignored the contribution from sources with higher space densities than the SWIRE 3.6 \micron\ sample.  The first assumption is 
difficult to model as it relies on an understanding of both the luminosity and clustering properties of the FIR emitting galaxies which are poorly constrained.  As we show in 
Table~\ref{tab:cirb} the second assumption is relative modest as we appear to resolve much of the FIR background within the SWIRE sample. However, we note that the SWIRE catalogues will not have resolved sources closer than the 3.6\micron\ beam and so will underestimate the number of close pairs.  Had such close pairs been included it would have counteracted the observed bias.    

We conclude from the simulations  that systematic effects due to source correlation in highly confused regions may underestimate the fluxes at low fluxes, but by a factor smaller than 0.5 dex..

\subsection{From stack fluxes to background intensities and luminosities}
The stacking technique naturally gives us the mean flux of galaxies in a class.  

To derive the contribution of that class to the background intensity ($\nu I_\nu$) we take the flux per unit frequency and divide by the unmasked area of the catalogue from which that class was selected i.e. $\nu I_\nu=\nu\sum{f_\nu}/\Omega$.

 The conversion to specific luminosity is more involved. To minimise differential effects across the cell and to include the appropriate completeness corrections we multiply each image in the stack (arising from galaxy $i$) by a scaling factor, $w_i$, before stacking (with $w_i=4\pi D_L(z_i)^2 K_{\rm FIR}(z_i) V_{\rm max,i}^{-1}M_{*i}^{-1}$). Then, rather than dividing the total stack by the number of galaxies, we divide by $\sum V_{\rm max,i}^{-1}$. The luminosity distance, $D_L$, and  K-correction, $K_{\rm FIR}$, transform flux into FIR luminosity while the stellar mass, $M_{*i}$, converts to specific luminosity and the $V_{\rm max}$ terms correct for the incomplete sampling of the cell volume.

\begin{figure*}
\includegraphics[width=16cm]{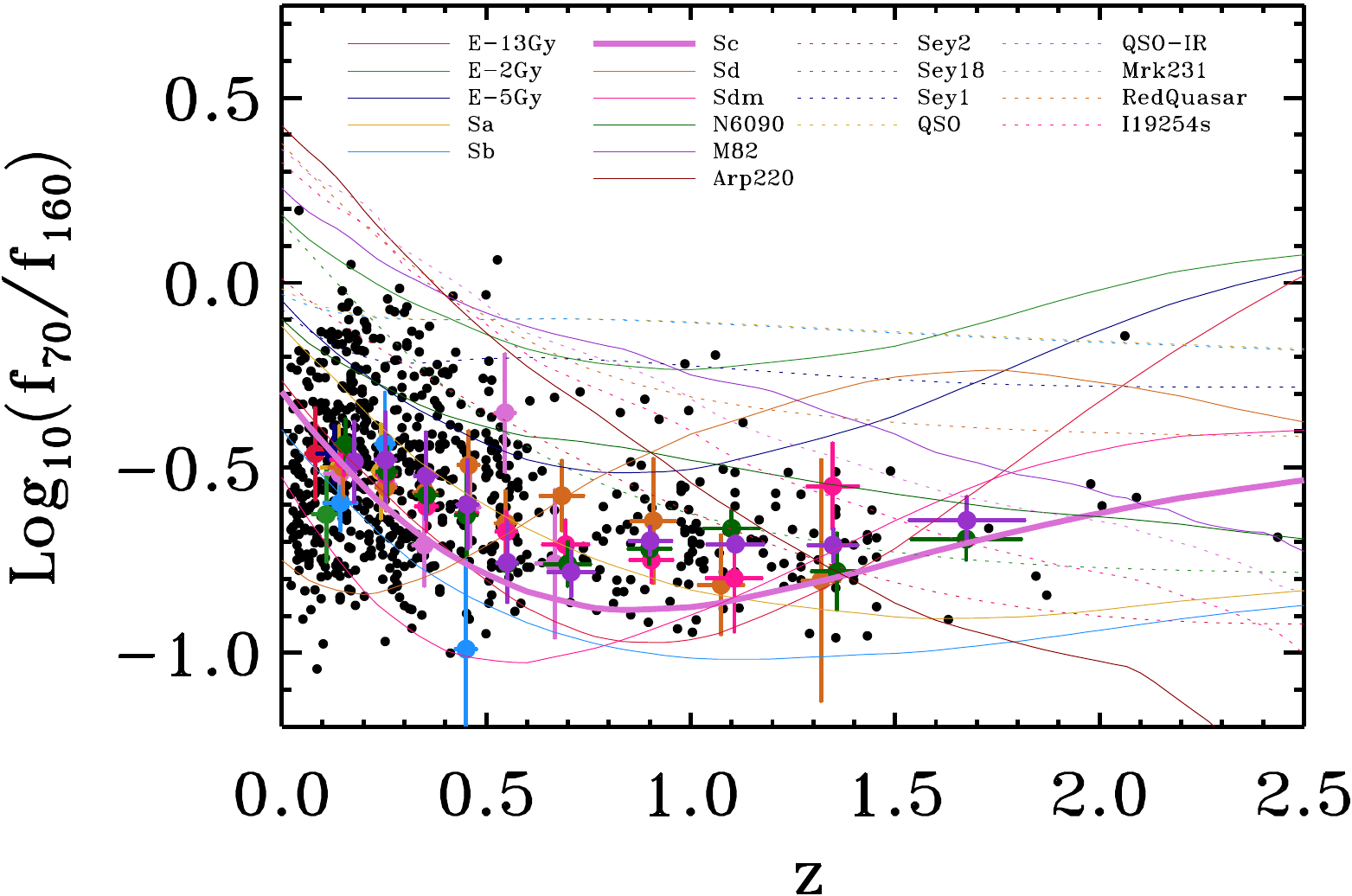}
\caption{$f_\nu$ colours ratios at 70/160 \micron\ for stack sub-samples. Filled circle data points are stack samples, colour-coded according to stellar mass using the same scheme as for Figures \protect\ref{fig:ssfrvsz}-
\protect\ref{fig:ssfvsz_red}. Black data points are catalogued galaxies extracted from the SWIRE Elais N1 field. Model templates are taken from \protect\cite{polletta07}.   }\label{fig:colours}
\end{figure*}

To calculate $K_{\rm FIR}$ we use a mean spectral energy distribution (SED) averaged over all galaxies in the cell. In Figure~\ref{fig:colours} we compare the 70/160 colours of all the cells with model templates from \citet{polletta07} and individual galaxies detected in both bands. The individual detections have similar colour distribution to the stack samples but with a tail to warmer colours.  This tail is expected because the detection criteria selects the most FIR luminous objects and there is a well known correlation between FIR luminosity and dust temperature (e.g. \citealt{chapman03}).   No single template fits all the samples at all redshifts. However, we see that the Sc template (plotted with a thicker line style) provides as good a fit as any other template over the range of classes and redshifts. This is natural as we'd expect the mean SED to be roughly the same as a galaxy with a moderate level of FIR activity. 

\begin{figure*}
\includegraphics[width=16cm]{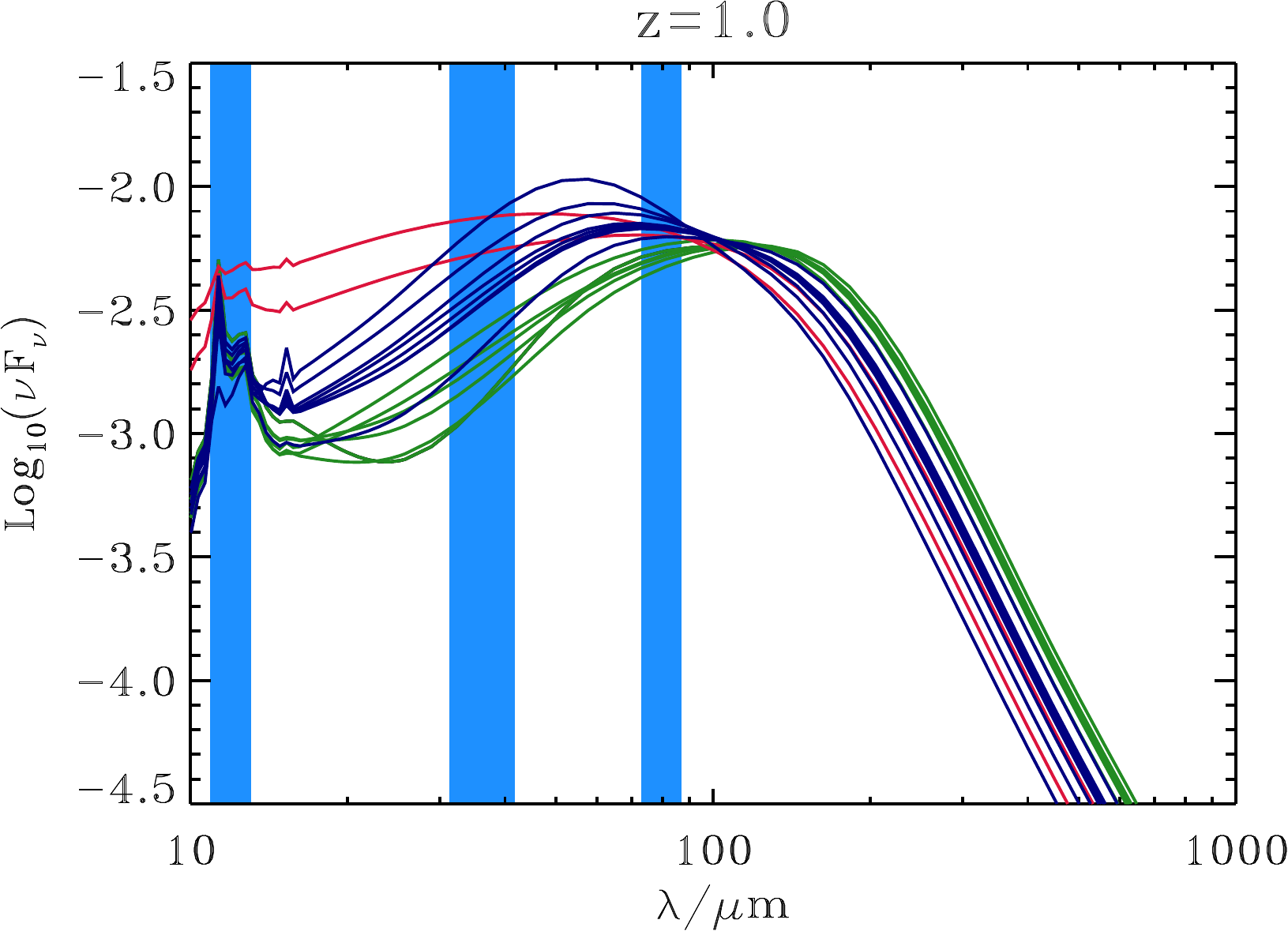}
\caption{Bolometrically normalised spectral energy distributions (SEDS) for a range of models from \protect\cite{xu}.  SEDs are shown in the rest-frame and are normalised to have the same power over the range $5<\lambda/\micron<1000$. The models include starburst galaxies, normal spirals and AGN. Overplotted are the 3 Spitzer MIPS bands (24, 70 \& 160\micron) for a galaxy observed at $z=1$. Notice that the bolometric normalisation is similar to normalisation at 100\micron.  Our analysis uses 70 and 160\micron, much previous work uses 24\micron. }\label{fig:bolsed}
\end{figure*}

We thus adopt an Sc template as a reasonable compromise between the different SEDs and use this 
to compute the FIR luminosity, $L_{\rm FIR}$, where we define the FIR range to be from $5\micron\le\lambda \le 1000\micron$, from 70 or 160 \micron\  fluxes.  
It should be noted that the Sc template peaks at 100\micron\ (in $\nu f_\nu$) or 130\micron\ (in $f\nu$) so the K-corrections at the two different wavelengths are different (in opposite directions for $z<0.2$).  Figure \ref{fig:bolsed} shows a wide variety of empirically based spectral energy distributions for galaxies detected in infrared bands.  we see that the 160\micron\ band on its own is reasonable bolometric power indicator, i.e. any uncertainties in which template is appropriate have a small effect on the K-correction.

If the far infrared luminosity traces the star-formation activity then the specific FIR luminosity is a measure of the specific star formation rate. However, the FIR luminosity also traces emission from AGN and from diffuse dust heated by the ambient stellar radiation field (``cirrus").  The AGN emission is warmer than either star-formation or cirrus and is expected to be less significant at these long wavelengths. This is confirmed by our SED analysis in Section~\ref{sec:cirb}.  The well-known FIR-Radio correlation (e.g. \citealt{Condon92}) suggests that even the cool ``cirrus" emission indirectly traces star-formation.  Strong AGN with modest extinction are excluded by the optical SED modeling.  Weaker or more obscured AGN will remain and provide some level of contamination. The level of this contamination is hard to estimate but typically AGN fractions are found to be $\sim30$\% (e.g. \citealt{polletta06}).  A strong AGN contamination would reveal itself in differences between our estimates arising from 70 and 160 \micron.  
Thus it is plausible to relate the total FIR luminosity to the obscured star formation rate. We relate the FIR luminosity 
to the total star formation rates (i.e. including an estimate of the unobscured contribution) using the conversion from \cite{rr97} and \cite{RowanRobinson07} expressed as:
$$ \frac{L_{\rm FIR}}{L_\odot}=0.51\times10^{10}\frac{\rm SFR}{M_\odot\, {\rm yr}^{-1}}$$ 
where we have taken the fraction of UV energy absorbed by dust to be $\epsilon=2/3$, 
and $L_{\rm FIR}/L_{\rm 60}=1.67$. These values are appropriate for an M82 like spectrum.  It is worth noting that if the obscuration is less than this, as you might expect for systems with less active star formation, then we will underestimate the star formation rate.

\subsection{Error estimation}\label{sec:errors}
To estimate the systematic uncertainties we calculate our errors using the variation in mean fluxes we get from field to field.  This approach accounts for sampling variance errors and some of the systematic errors arising from our use of photometric redshifts.
Although our photometric redshift technique is the same across all the fields, the optical data comes from different telescopes; this means the field to field variations
could be significant if the photo-$z$ have any subtle dependencies on the  bands or optical limits.  

The areas of each field $\Omega_i$ are given in Section~\ref{sec:masks}. 
For each sub-sample we calculate an average flux (or specific star-formation) weighted by these areas
$\bar{f}=\sum_i \Omega_i\, f_i/\sum_i \Omega_i$.  We estimate the error on the resulting weighted average as 

$$ \sigma_{\bar{f}}^2\ = \frac{1}{\sum_{i}}\frac{ \sum_{i}{\Omega_i} }{ \left(\sum_{i}{\Omega_i}\right)^2 - \sum_{i}{{\Omega_i}^2} }\ 
\sum_{i}{{\Omega_i}\left(f_i - \bar{f}\right)^2}
 $$

The first term is a factor to scale from the population variance to the variance in the mean, the next two terms are the weighted estimator for 
the population variance.\footnote{e.g. http://pygsl.sourceforge.net/reference/pygsl/node36.html}

Any stacks with fewer than 9 galaxies (the threshold for reasonable $\chi^2$ found in Section \ref{sec:method}) were excluded from the average and any cells that had fewer than 100 galaxies after the averaging were excluded.

\section{Results}

\subsection{Contributions to CIRB from various observational sub-samples}\label{sec:cirb}
\begin{table*}
\input{table1.tex}

\caption{Contributions to the 70\micron\ and 160\micron\ FIR background from various populations.  All catalogues and maps have been 
masked by the mask described in the text. The first four rows are galaxies extracted from the Spitzer data only catalogues. 
The fifth row comes from the Spitzer/optical cross-matched catalogues. The remaining rows above the line are extracted from the photo-$z$ catalogues of \protect\cite{RowanRobinson07} which has been filtered to remove galaxies with poor quality photo-$z$ and classifications (i.e. those with $n_{\rm band}<4,$ or  
$\chi^2\ge5$). The data are averaged over four fields Elais-N1, Elais-N2, Lockman, CDFS 
weighted by the unmasked areas of 4.33, 2.41, 2.75, 1.85  sq. deg. respectively for a total of 11.33 sq. deg.  Errors are deduced from the field to field variations.  
The last two rows are estimates extracted from Table~1 of \protect\cite{dol06}}\label{tab:cirb}
\end{table*}

Before we examine the specific far infrared luminosity we explore the contribution of different 
subsets of the SWIRE catalogues to the cosmic infrared background (CIRB).  These are tabulated in Table~\ref{tab:cirb}.  

The SWIRE galaxies which are bright enough to be detected individually at 70 or 160\micron\ (rows ~3~\&~4) contain only a small fraction ($\simlt 12$\%) of the CIRB information of the 
SWIRE data set as a whole (row 1) which provides strong motivation for our stacking analysis.  The optical sub-sample (row 5) contains about 60\% of the FIR information. This selection is similar to that used in most of this paper. It is encouraging that we detect this much of the FIR flux, however, it emphasises the need for deeper optical data to fully exploit the SWIRE data.  The SWIRE 24 \micron\ catalogues detect about half of the CIRB seen in the 3.6\micron\ sources.  We find that SWIRE 24\micron\ catalogues resolve 30-40\% of the CIRB seen by \citet{dol06} in fainter 24\micron\ samples, while the SWIRE 3.6\micron\ catalogues resolve 70-80\%.  Estimation of the total CIRB from direct measurement is highly uncertain but if we use the estimates quoted by \citet{dol06} then the SWIRE 3.6\micron\ catalogues resolve about 70\% of the 70\micron\ background and 60\% of the 160\micron\ background.  Roughly half of the background is resolved in the optical samples at $r<23.5$.

Some other points to note from this table are that the FIR colours generally become warmer as we move from early to late types, starbursts and AGN (as expected).  As we noted before when discussing Figure~\ref{fig:colours},  the colours of sources catalogued at 70 or 160\micron\ are warmer than those selected at other bands.  This is because these sources tend to have higher FIR luminosity and are thus preferentially star-burst galaxies or AGN. This effect is less pronounced at 160\micron\ which picks up cooler sources.

\subsection{Specific far-infrared luminosity as a function of stellar mass}\label{sec:ssfr}
To explore the specific star-formation  we measure the ratio of far-infrared luminosity to stellar-mass i.e. the specific FIR luminosity. An early exploration of this as a function of optical luminosity (as a proxy for stellar mass) from ISO data was presented by \cite{oliverpozzi}.  We use the stellar mass and redshift cells described in Section~\ref{sec:bin}.

\begin{figure*}
\includegraphics[width=16cm]{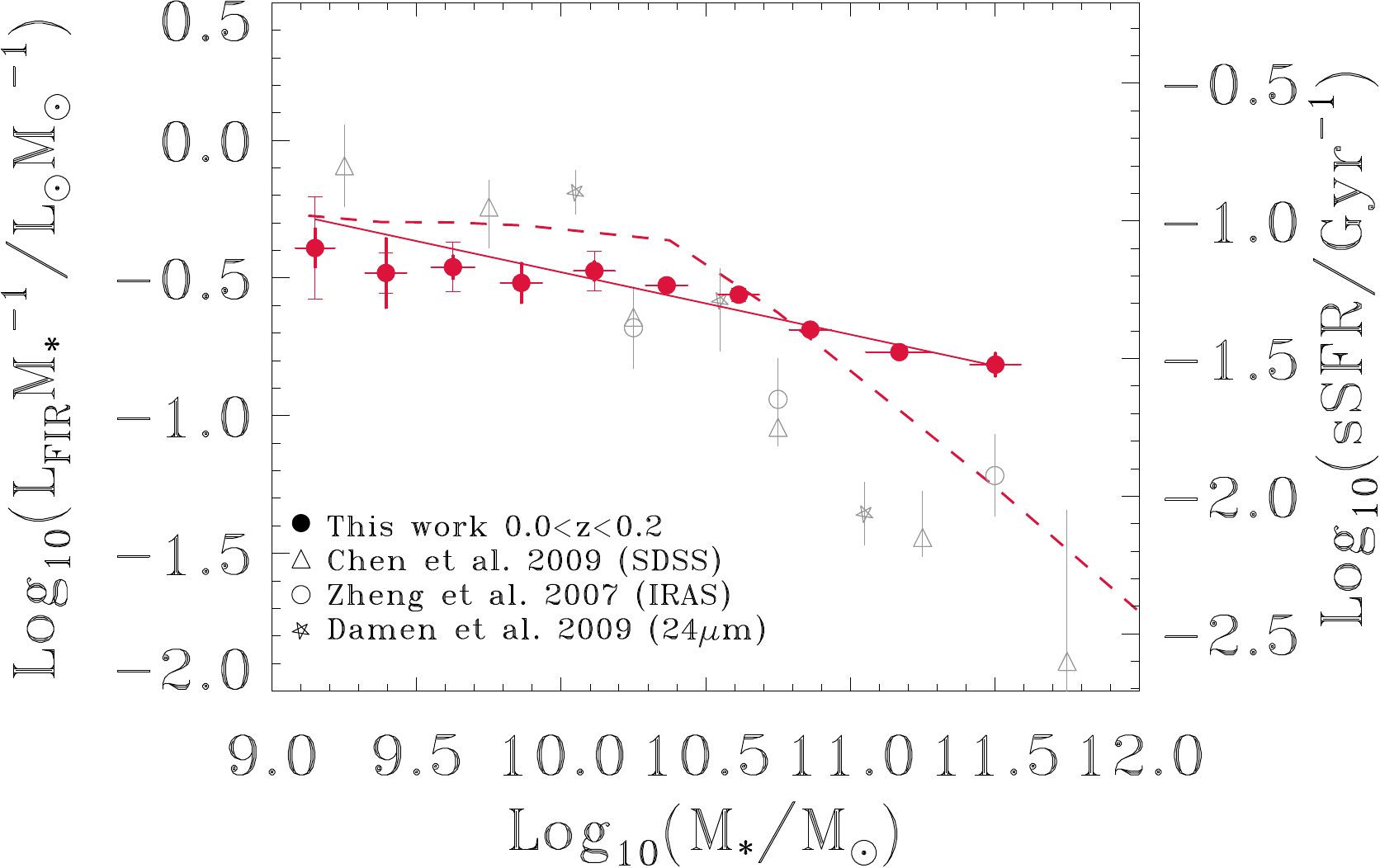}
\caption{Specific FIR luminosity or specific star-formation rate as a function of stellar mass in redshift classes. Redshift ranges from $0<z<0.2$.  Filled circles are the average of our estimates from the 70 \&160\micron\ data which are the weighted average over four fields (Elais N1, Elais N2, Lockman and CDFS).   Thick vertical  error bars come from the field-to-field scatter and thin error bars with hats from the variation between 70 \& 160\micron\ estimates. All points have used at least 100 galaxies in the stacking analysis.  Solid lines are our power-law trends with parameters give in Table~\protect\ref{tab:massfits}.  Calibration to specific star-formation (right-had axis) is given in the text. Previous estimates of specific star-formations rates are estimated from 24 \micron\ data (stars) \protect\cite{damen09} ($0.1<z<0.3$), IRAS (open circles) \protect\cite{zhe07} $z\simlt0.02$, and from SDSS spectra (triangles) $z\sim0.1$ \protect\cite{chen09}. Dashed lines are predictions from the semi-analytic models of \protect\cite{Delucia2007}. 
}\label{fig:ssfvsmass0}
\end{figure*}

\begin{figure*}
\includegraphics[width=16cm]{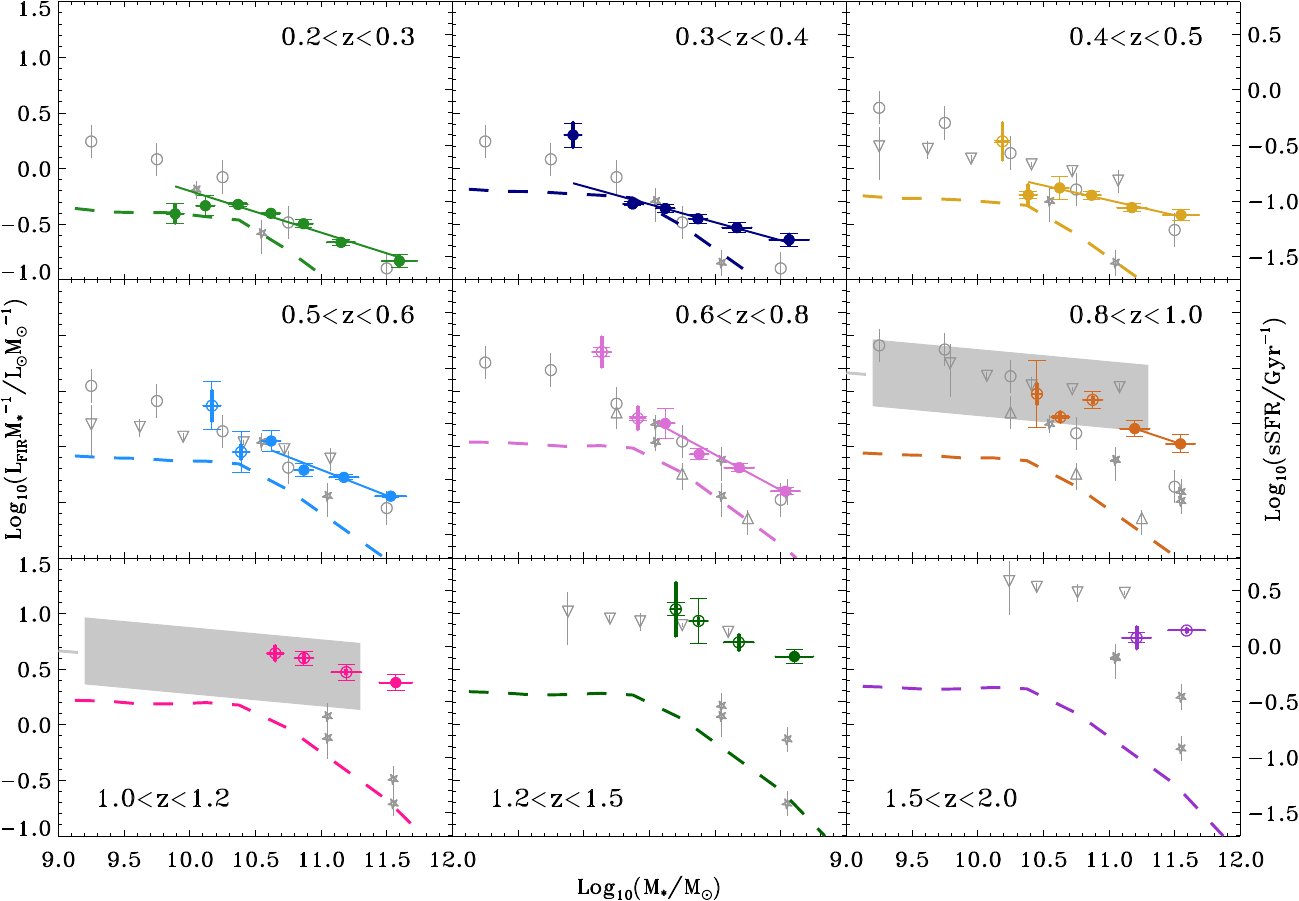}
\caption{Specific FIR luminosity or specific star-formation rate as a function of stellar mass in redshift classes.  Mean points are the weighted average over four fields (Elais N1, Elais N2, Lockman and CDFS) and over 70 \&160\micron\ data.  Error bars are estimated from the scatter between fields and bands as in Figure~\protect\ref{fig:ssfvsmass0}.  Incomplete cells where not all galaxy templates are represented are indicated with open circles.  Power law fits to complete data are shown with solid lines. Comparison samples from similar redshift ranges are plotted with fainter symbols. Since the redshift bins are not identical these sometime appear on more than one plot or with data from two bins in the original work on the same plot.  Estimates for combined  24\micron\ and GALEX UV photometry come from 
\protect\cite{elb07} (shaded, $0.8<z<1.3$)
\protect\cite{zhe07} (circles, $z$ bins, 0.2-0.4, 0.4-0.6, 0.6-0.8,0.8-1.0)  and \protect\cite{damen09}  (stars, $z$ bins 0.1-0.3, 0.3-0.5, 0.5-0.7, 0.7-0.9, 0.9-1.1, 1.1-1.3, 1.3-1.5, 1.5-1.7, 1.7-1.9 ). Radio estimates from \protect\cite{dunne09} (down triangles, $z$ bins 0.2-0.7, 0.7-1.2, 1.2-1.7, 1.7-2.2)  and from DEEP 2 spectra \protect\cite{chen09}  (up triangles, $z\sim1$). The parameters of the power-law fits to are data are tabulated in Table \protect\ref{tab:massfits}. Dashed lines are predictions from the semi-analytic models of \protect\cite{Delucia2007}
}\label{fig:ssfvsmass}
\end{figure*}

 In Figures~\ref{fig:ssfvsmass0} and \ref{fig:ssfvsmass} we plot the specific FIR luminosity as a function of stellar mass. We omit points where the fractional error (calculated using the field-to-field variations) is more than 1.   Data points flagged as incomplete (Section \ref{sec:selectioneffects}) are unfilled.

We see similar results at both 70 and 160 \micron. This agreement is an important validation of our method as the data are completely  independent and the K-corrections applied are very different.  We are thus confident in combining the two data sets to give a single estimate of the specific FIR luminosity (being a weighted average of the estimate at each wavelength). The scatter between the two independent wavelengths is indicated in the error bars.

In Figure~\ref{fig:ssfvsmass0} we compare our average specific FIR-luminosity in our lowest redshift bin (where we cover the widest mass range) with other data from the literature. We see a simple power-law trend but a shallower slope, particularly with higher specific star-formations at higher masses.  The \cite{chen09} work models the SDSS spectra and is sensitive to star-formation over a longer time-scale ($\sim 1 Gyr$) than the FIR.  This may lead them to find higher specific star formation rates due to the evolution over that time-scale.  Alternatively, this may be an obscuration effect, if the optical observations miss heavily obscured star-formation and if that deficit is greater in higher mass systems.  This might be expected if star formation in massive systems tends to be in more deeply embedded sites as is seen in Arp 220.   The analysis of \citet{damen09} includes star formation measures from 24 \micron\ and so should include obscured star formation, but may miss cooler contributions that we pick up at longer wavelengths.  This latter explanation is tentatively supported by the IRAS measurements  from \citet{zhe07b} at longer wavelengths which lies between our results and those of  \citet{damen09}.  The remaining discrepancy between our work and \citet{zhe07b} (at lower $z$) could then be explained as due to evolution across our redshift bin, which will preferentially increase the sSFR in the higher mass bins.  A final possibility is that there are inconsistencies in the stellar mass estimates, either through modelling or through photo-$z$ estimates. This is likely to be more of a problem for us at these lower $z$, where photo-$z$ errors are more significant, than it is at higher $z$.

Comparing with data at higher redshifts (Figure \ref{fig:ssfvsmass}) the agreements are much better.  Our large area means we are able to divide our data into finer redshift bins than previous work while still maintaining high statistical precision. We find similar slope and amplitude to obscured tracers e.g. estimated from 24 \micron\ data \citep{zhe07b}.  Their observations were over a much smaller field and so they were limited to lower stellar masses.   At  $z>1$ we find sSFR significantly higher than \citet{damen09} (who included FIR indicators), but similar to those found by \citet{dunne09} (who use radio).  Our slope is hard to compare with \citet{dunne09} as their redshift ranges are much larger and differential evolution across their bin will be significant.
\citet{elb07}  derived an empirical relation from 24\micron\ and UV data and quote ${\rm SFR} /(M_\odot{\rm yr}^{-1})=7.2 [M_*/(10^{10}M_\odot)]^{0.9}$ at $0.8<z<1.2$, again limited to lower stellar masses.   The \cite{elb07} measurement  has a consistent amplitude in the limited range where our stellar mass ranges overlap. However, they have a shallower slope than ours and that of \citet{zhe07b}; this might be an effect of differential evolution across their wide redshift bin.
 
Overall it seems that the relations are smooth and there is no strong evidence for any breaks.   However, there is a need for consistent measurements over a wide dynamic range in mass and in narrow redshift bins.  

There is a strong evolutionary trend which we explore in more detail in Section~\ref{sec:evol}.  Looking within each redshift bin (while bearing in mind that we only see lower mass objects at lower redshifts) we see that there is a consistent decline in specific FIR luminosity as we move to higher stellar mass. 
We model this trend as 
$${\rm sSFR}\equiv\frac{{\rm SFR}}{M_{\rm *}}\propto\left(\frac{M}{10^{11}M_\odot}\right)^\beta.$$
This power-law model is a reasonable fit and parameters for the average fit are given in Table~\ref{tab:massfits}.  There is some variation in the slope with redshift, the slope is steeper for $0.6<z<0.8$, but a mean value of $\beta=-0.38\pm0.14$ provides a plausible description of the data.

For a direct comparison with semi-analytic models we take the simulations of \citet{Delucia2007}\footnote{Publicly available on the Millennium Simulation data download site (see \citealt{Lemson2006}).}
 and select galaxies in the same stellar mass and redshift cells as our data. This model includes AGN feedback which effectively suppresses star formation for galaxies with large central black holes (typically galaxies with M$>10^{10.5}$M$_{\odot}$).  The comparison is shown in Figure~\ref{fig:ssfvsmass0} for our lowest $z$ bin and in Figure~\ref{fig:ssfvsmass} for the others. This model fits our low mass data well, but the model under-predicts our high-mass end.  The same model agrees better with other data sets at the high mass end but not as well at lower mass.    The shape of the model curves are identical at higher redshifts with just the normalisation increasing.  The clear break in the model curves arises at the mass scale where the AGN feedback becomes important.   We don't see such a break in any redshift bin.   In addition,   this model fails to predict the amplitude of the evolution with redshift which we discuss in Section \ref{sec:evol}.

\begin{figure*}
\includegraphics[width=16cm]{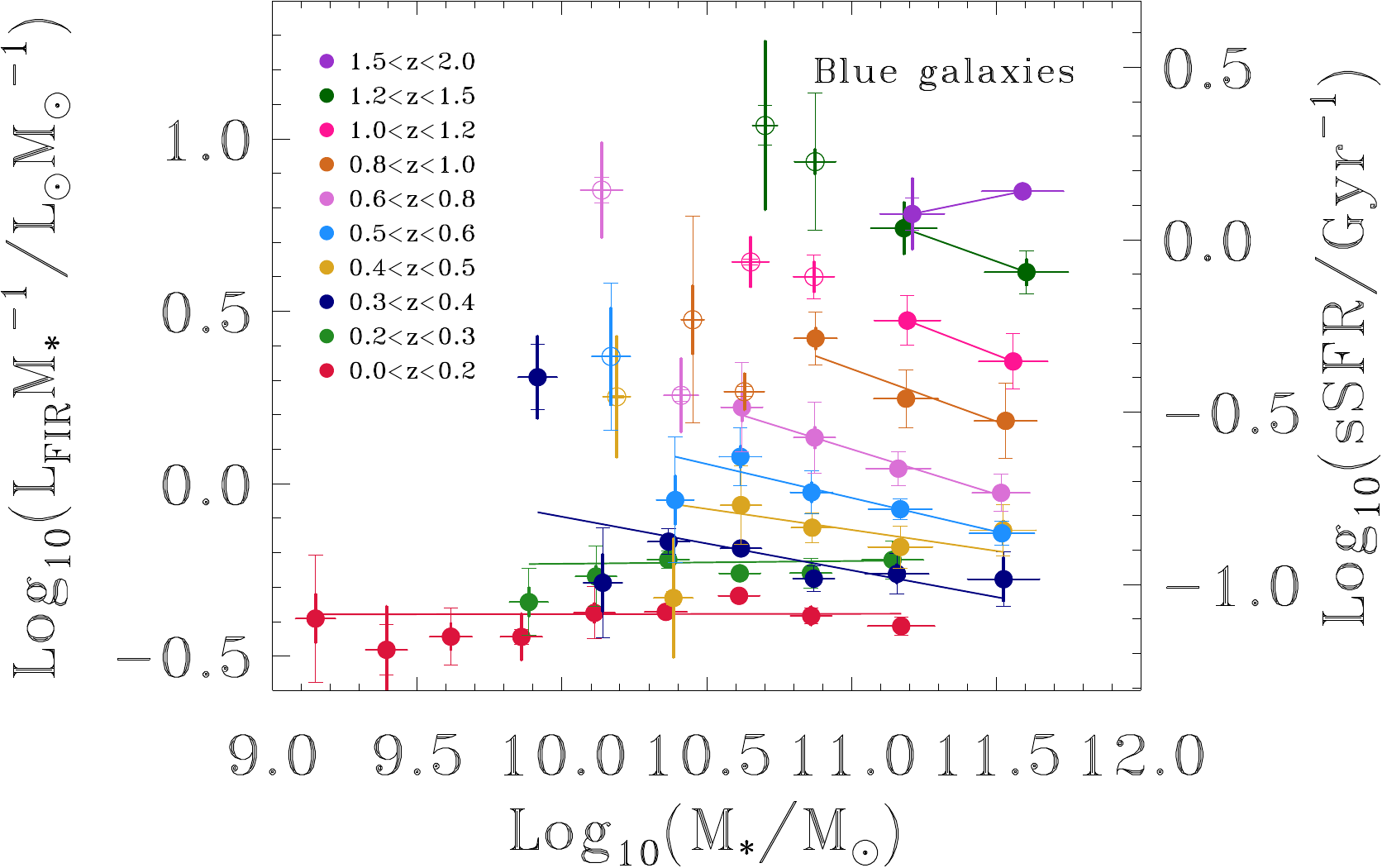}
\caption{Specific FIR luminosity or specific star-formation rate as a function of stellar mass in redshift classes  for ``blue" galaxies, i.e. those with SED types 3-11. Points are the weighted average over four fields (Elais N1, Elais N2, Lockman and CDFS) and over 70 \&160\micron\ data with error bars are as for Figures \protect\ref{fig:ssfvsmass0} and \protect\ref{fig:ssfvsmass}. Incomplete points are unfilled.  The parameters of the power-law fits to are data are tabulated in Table \protect\ref{tab:massfits}. }\label{fig:ssfvsmass_blue}
\end{figure*}

\begin{figure*}
\includegraphics[width=16cm]{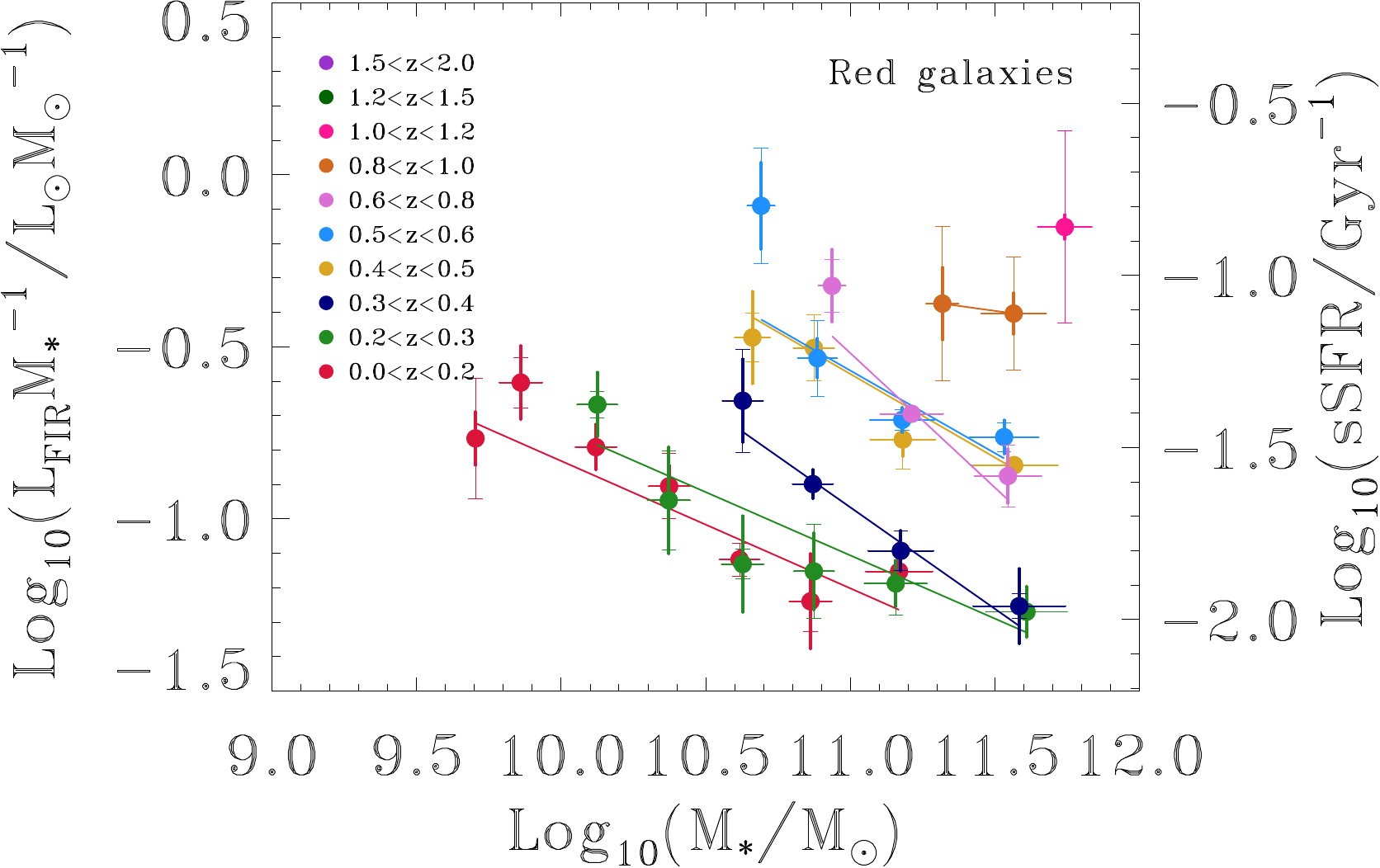}
\caption{Specific FIR luminosity or specific star-formation rate as a function of stellar mass in redshift classes  for ``red" galaxies, i.e. those with SED types 1-2. Points are the weighted average over four fields (Elais N1, Elais N2, Lockman and CDFS) and over 70 \&160\micron\ data with error bars are as for Figures \protect\ref{fig:ssfvsmass0} and \protect\ref{fig:ssfvsmass}. Incomplete points are unfilled. The parameters of the power-law fits to are data are tabulated in Table \protect\ref{tab:massfits}. }\label{fig:ssfvsmass_red}
\end{figure*}

We perform the same analysis for galaxies separated by the optical/NIR SED class.  We show the separation into late-type galaxies (SED types 3-11) in Figure \ref{fig:ssfvsmass_blue}  and early-type galaxies (SED types 1-2, Figure \ref{fig:ssfvsmass_red}) with fit parameters in Table \ref{tab:massfits}.  It is immediately striking that the relation ships are much steeper (though naturally lower amplitude) for the early type galaxies $\beta\sim-0.46$ q.v. $\beta\sim-0.15$ for the late-type galaxies.  This may indicate that early type galaxies harbour some components of star-formation that are unrelated to the dominant host galaxy (e.g. through accretion of gas rich  companions).  However, this cannot be the complete picture as totally uncorrelated star-formation onto  dead hosts would produce a steeper slope with $\beta=-1$.

We have performed the same analysis on all of the different optical galaxy classifications and these show a logical progression between the behaviours seen in the early and late groupings shown here.  A figure illustrating this is shown in \citet{roseboom09}.

If we ignore the highest redshift bin, for which the slope is poorly constrained, there is some indication from these figures and Table  \ref{tab:massfits} that the slope of the relation between sSFR and stellar mass becomes steeper with increasing redshift for both early and late type galaxies and for all galaxies combined.

\begin{table*}
\begin{tabular}{ccccccccccc}
\hline
\multicolumn{2}{c}{} & 
\multicolumn{3}{c}{All galaxies} & 
\multicolumn{3}{c}{Blue galaxies} & 
\multicolumn{3}{c}{Red galaxies} \\
$z_{\rm min}$ & $z_{\rm max}$ & $\log_{10}Y$ & $\beta$ & $\chi^2/\nu$& $\log_{10}Y$ & $\beta$ & $\chi^2/\nu$& $\log_{10}Y$ & $\beta$ & $\chi^2/\nu$\\
& & $\log({\rm Gyr}^{-1})$ &&&$\log({\rm Gyr}^{-1})$ &&&$\log({\rm Gyr}^{-1})$  \\
\hline
0.0	&	0.2	&	-1.41	$\pm$	0.01	&	-0.23	$\pm$	0.02	&	2.28	&	-1.08	$\pm$	0.01	&	0.00	$\pm$	0.02	&	2.26	&	-1.91	$\pm$	0.03	&	-0.37	$\pm$	0.06	&	2.73	\\
0.2	&	0.3	&	-1.28	$\pm$	0.01	&	-0.37	$\pm$	0.03	&	3.48	&	-0.93	$\pm$	0.01	&	0.01	$\pm$	0.01	&	3.58	&	-1.81	$\pm$	0.04	&	-0.37	$\pm$	0.08	&	1.10	\\
0.3	&	0.4	&	-1.19	$\pm$	0.02	&	-0.32	$\pm$	0.04	&	3.91	&	-0.96	$\pm$	0.01	&	-0.16	$\pm$	0.03	&	3.99	&	-1.67	$\pm$	0.03	&	-0.59	$\pm$	0.14	&	0.50	\\
0.4	&	0.5	&	-0.99	$\pm$	0.02	&	-0.27	$\pm$	0.05	&	0.71	&	-0.84	$\pm$	0.01	&	-0.12	$\pm$	0.03	&	4.12	&	-1.29	$\pm$	0.02	&	-0.49	$\pm$	0.05	&	2.28	\\
0.5	&	0.6	&	-0.91	$\pm$	0.02	&	-0.46	$\pm$	0.05	&	4.65	&	-0.75	$\pm$	0.01	&	-0.20	$\pm$	0.04	&	1.63	&	-1.28	$\pm$	0.03	&	-0.48	$\pm$	0.11	&	5.11	\\
0.6	&	0.8	&	-0.78	$\pm$	0.01	&	-0.64	$\pm$	0.06	&	1.66	&	-0.61	$\pm$	0.02	&	-0.26	$\pm$	0.04	&	0.45	&	-1.23	$\pm$	0.05	&	-0.78	$\pm$	0.21	&	2.78	\\
0.8	&	1.0	&	-0.47	$\pm$	0.03	&	-0.39	$\pm$	0.07	&		&	-0.37	$\pm$	0.02	&	-0.30	$\pm$	0.05	&	5.33	&	-1.04	$\pm$	0.26	&	-0.12	$\pm$	0.50	&		\\
1.0	&	1.25	&				&				&		&	-0.17	$\pm$	0.04	&	-0.32	$\pm$	0.07	&		&				&				&		\\
1.25	&	1.5	&				&				&		&	0.09	$\pm$	0.11	&	-0.30	$\pm$	0.20	&		&				&				&		\\
1.5	&	2.0	&				&				&		&	0.04	$\pm$	0.16	&	0.17	$\pm$	0.28	&		&				&				&		\\ 
\hline

	&		&				&	-0.38	$\pm$	0.14	&		&				&	-0.15	$\pm$	0.16	&		&				&	-0.46	$\pm$	0.21	&		\\
\end{tabular}
\caption{Power-law fits to the specific star-formation as a function of stellar mass as plotted in Figures \protect\ref{fig:ssfvsmass0}-\protect\ref{fig:ssfvsmass_red}.  
Modelled as $sSFR=Y(M/10^{11}M_\odot)^\beta$ in redshift ranges indicated. Combined 70 and 160 \micron\ data. Averages and standard deviations of $\beta$ are given below the line.  Fits are calculated for all galaxies and separately for ``blue" (templates Sab-Sdm and starburst, $3\le j_2\le11$)  and ``red" galaxies (both E templates, $j_2\le2$). Reduced $\chi^2$ are quoted if the number of points used in the fit is more than 2.}\label{tab:massfits}
\end{table*}

\subsection{Specific FIR luminosity as a function of redshift}\label{sec:evol}
%
\begin{figure*}
\includegraphics[width=16cm]{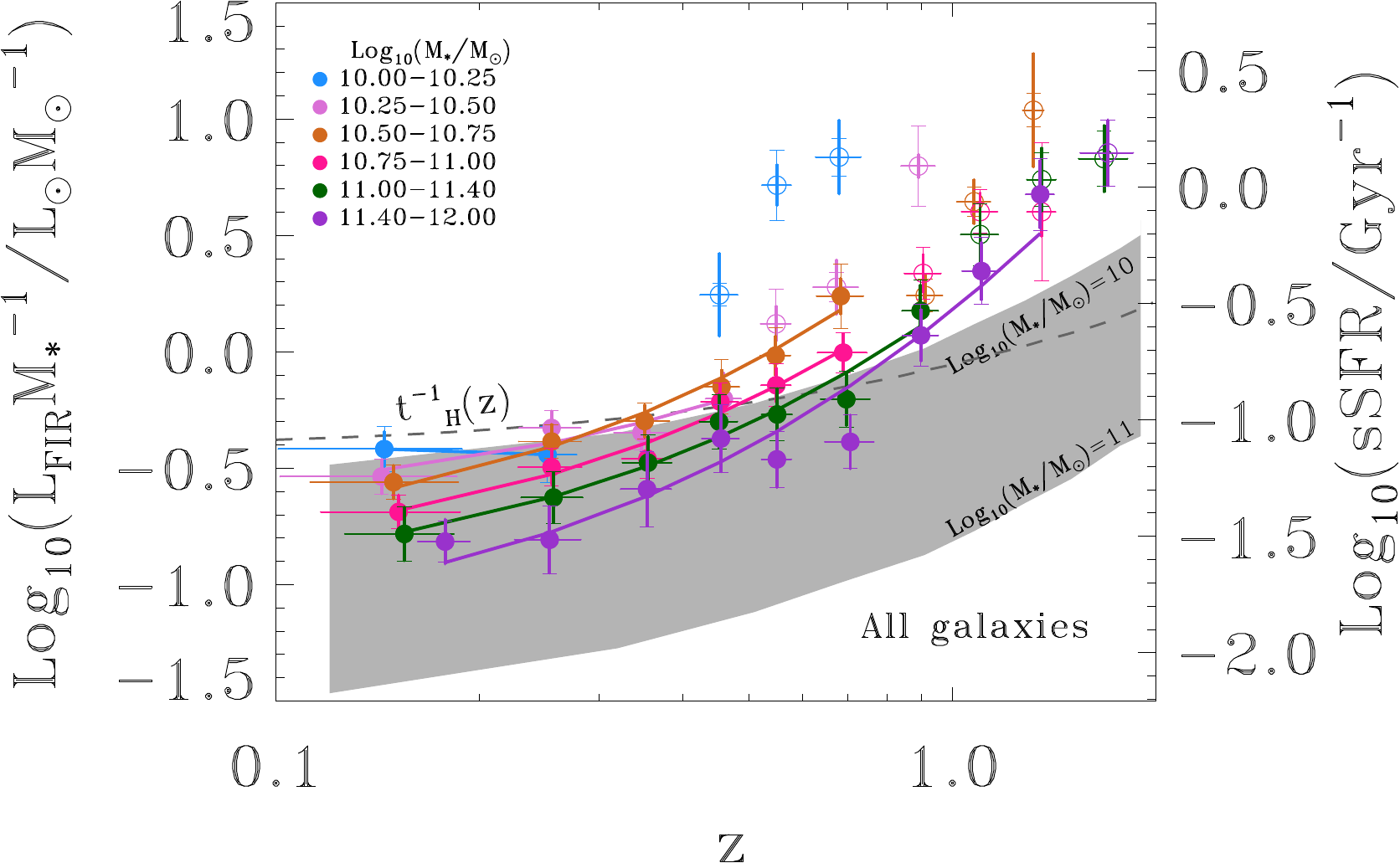}
\caption{Specific FIR luminosity or specific star-formation rate as 
a function of redshift in stellar mass classes. Filled circles are the average of our estimates from the 70 \&160\micron\ data which are the weighted average over four fields (Elais N1, Elais N2, Lockman and CDFS).   Thick vertical  error bars come from the field-to-field scatter and thin error bars with hats from the variation between 70 \& 160\micron\ estimates. Solid lines are our power-law trends with parameters give in Table~\protect\ref{tab:zfits}.   The shaded region represent the semi-analytic models of \protect\cite{Delucia2007} with the upper and lower bounds being the predictions for $M_*=10^{10}M_\odot$ and $10^{11}M_\odot$.  The dashed line shows the inverse Hubble time, galaxies above this line are `bursting' producing stars at a higher rate than their historical average. }\label{fig:ssfrvsz}
\end{figure*}
\begin{figure*}
\includegraphics[width=16cm]{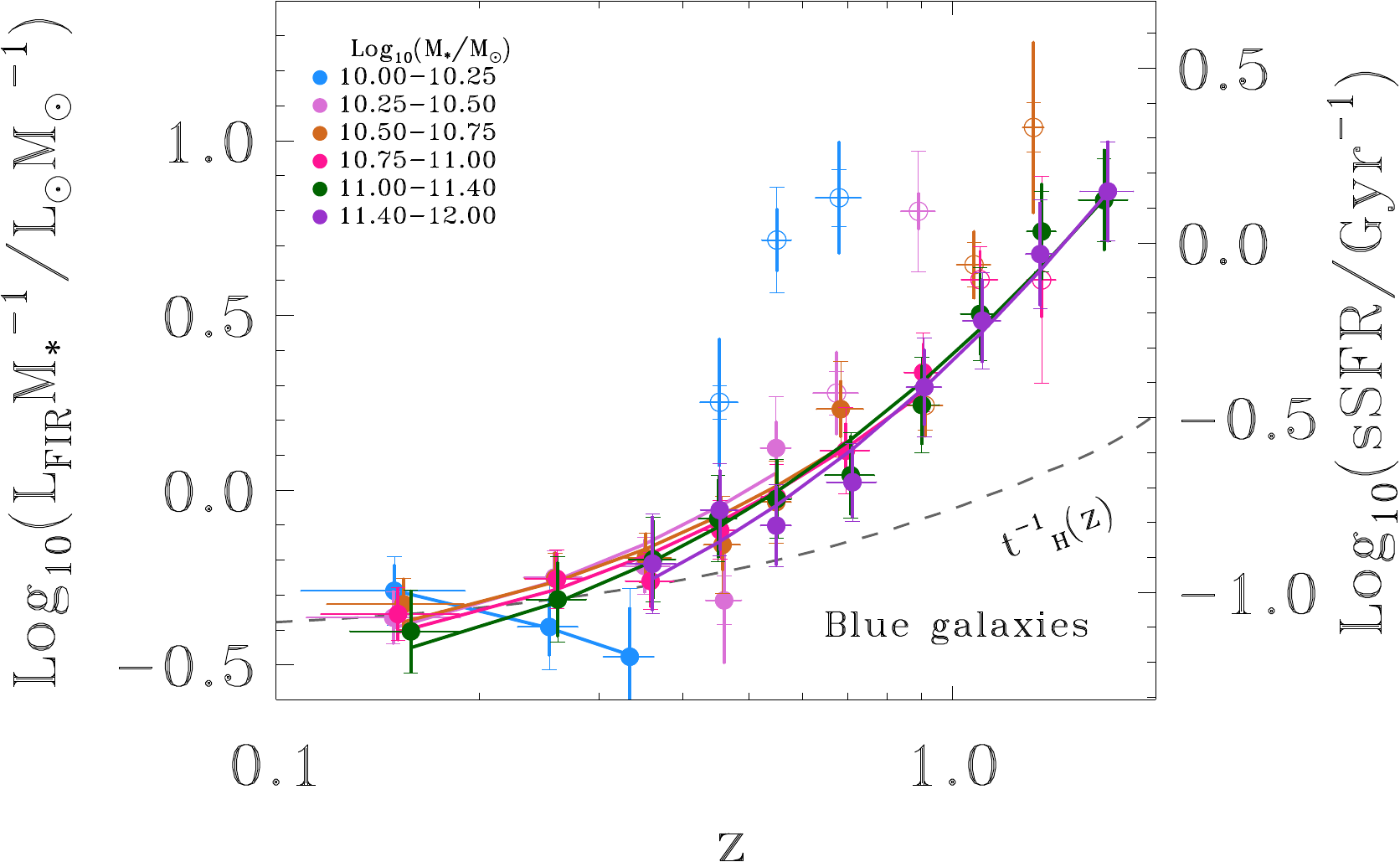}
\caption{Specific FIR luminosity or specific star-formation rate as 
a function of redshift in stellar mass classes for ``blue" galaxies, i.e. those with SED types 3-11.   Filled circles are the average of our estimates from the 70 \&160\micron\ data which are the weighted average over four fields (Elais N1, Elais N2, Lockman and CDFS).   Thick vertical  error bars come from the field-to-field scatter and thin error bars with hats from the variation between 70 \& 160\micron\ estimates. Solid lines are our power-law trends with parameters give in Table~\protect\ref{tab:zfits}.  The dashed line shows the inverse Hubble time, galaxies above this line are `bursting' producing stars at a higher rate than their historical average. }\label{fig:ssfvsz_blue}
\end{figure*}
\begin{figure*}
\includegraphics[width=16cm]{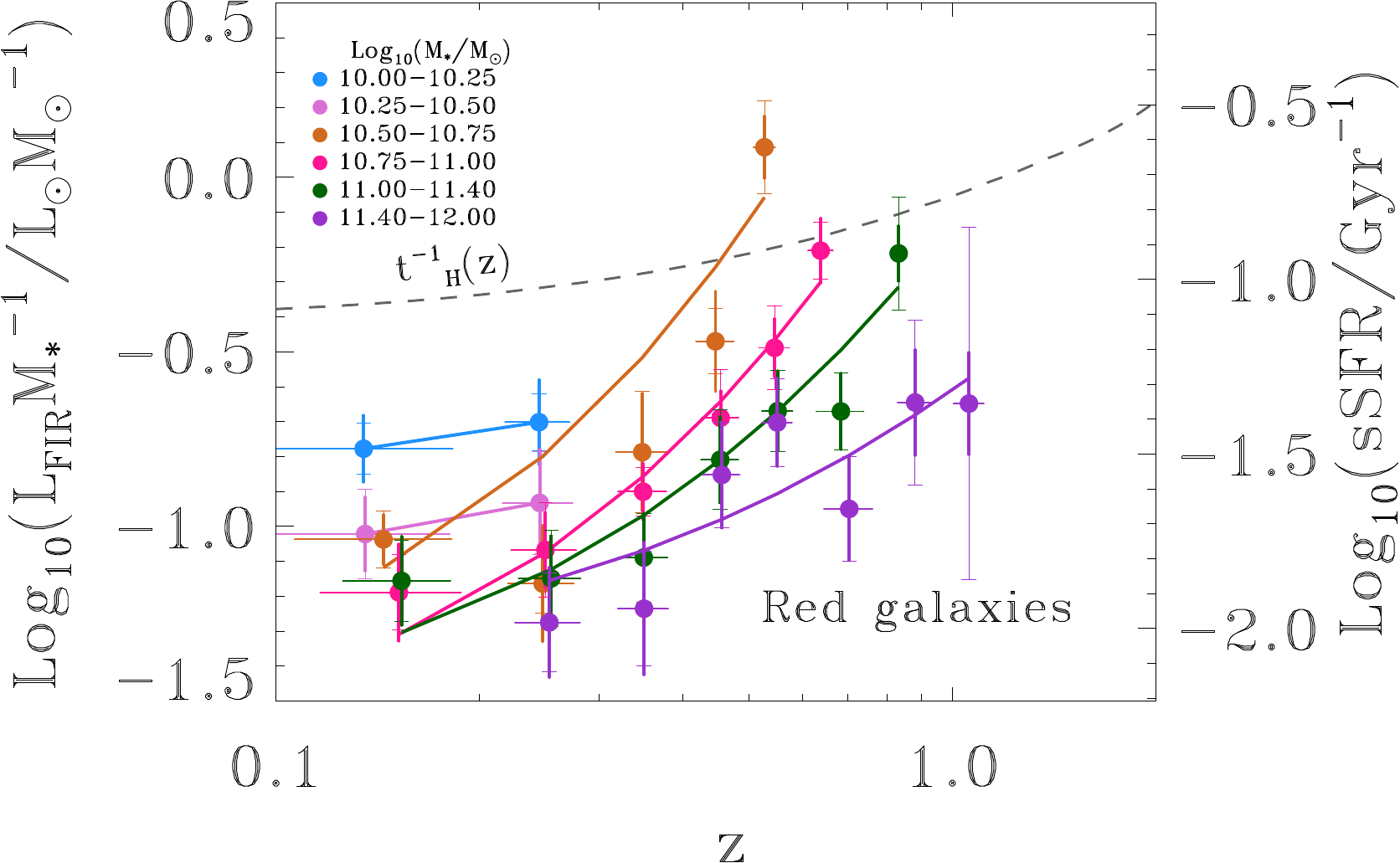}
\caption{Specific FIR luminosity as a function of redshift in stellar mass classes for ``red" galaxies, i.e. those with SED types 1-2. Filled circles are the average of our estimates from the 70 \&160\micron\ data which are the weighted average over four fields (Elais N1, Elais N2, Lockman and CDFS).   Thick vertical  error bars come from the field-to-field scatter and thin error bars with hats from the variation between 70 \& 160\micron\ estimates. Solid lines are our power-law trends with parameters give in Table~\protect\ref{tab:zfits}.   The dashed line shows the inverse Hubble time, galaxies above this line are `bursting' producing stars at a higher rate than their historical average. }\label{fig:ssfvsz_red}
\end{figure*}

By plotting the same data as a function of redshift but in stellar mass classes the evolutionary trend is readily apparent, see Figure~\ref{fig:ssfrvsz}.   It is immediately clear that there is a dramatic increase 
in FIR to optical ratio (or specific FIR luminosity) by a factor of $>100$ over the interval $0\le z \le 2.4$. This is seen at all masses apart from the lowest mass bin which has poor statistics and a limited redshift baseline. 
Again, we see the same trends in both the 70 and 160\micron\ data. This is an important corroboration as the FIR data are independent and the K-corrections will be different.We have fit this trend with a simple model
$${\rm sSFR}=X(1+z)^\alpha.$$
The best fits are shown in Figures~\ref{fig:ssfrvsz}, \ref{fig:ssfvsz_blue} and~\ref{fig:ssfvsz_red} and the parameters and
goodness of fit are given in Table~\ref{tab:zfits}. 

The simple power-law fit provides a good description of the data in most cases.  As already seen in Section \ref{sec:ssfr}, the zero redshift specific FIR luminosity, i.e. the parameter $X$, declines with increasing stellar mass. We see no strong indication that $\alpha$ increases with increasing mass (in the range $10^{10.5}M_\odot<M<10^{12}$), as would be expected in the ``down-sizing" scenario.

We also compare with the observations from \citet{zhe07b}.  Their data shows similar evolutionary behaviour at lower stellar masses.  If we assume that the variations of evolutionary rate
with stellar mass are statistical variations then we can make a simpler model with the same $\alpha$ for all classes.  We take the average of the $\alpha$ estimates in Table~\ref{tab:zfits} for $M>10^{10.5}M_\odot$.  From this we deduce that the 
 $L_{\rm FIR}/M_{\rm *}$ ratio varies as $(1+z)^{4.4\pm0.3}$, with the error bar being the standard deviation of the measurements.  The lowest mass bins are, however, inconsistent both with this mean evolutionary rate and the lower mass estimates from \citet{zhe07b}.
 
We show the redshift trend for the SAMs in Figure~\ref{fig:ssfrvsz}. 
The SAMs  show specific star formation rates decreasing smoothly with cosmic time as the gas supplies are declining and the stellar masses are building up.  The trend with redshift in the SAMs has a mean slope of $\alpha=2.4$.   
which is considerably shallower than the observed data  ($\alpha=4.4\pm0.3$).

\begin{table*}
\begin{tabular}{crrrrrrrrrrrrrr}
\hline
&  \multicolumn{3}{c}{All galaxies} & \multicolumn{3}{c}{``Blue'' galaxies}& \multicolumn{3}{c}{``Red'' galaxies}  \\
\multicolumn{1}{c}{$\log_{10}(M_{\rm *}/M_\odot)$} & 
\multicolumn{1}{c}{$\log_{10}{(X/{\rm Gyr^{-1}})}$} & 
\multicolumn{1}{c}{$\alpha$} & 
\multicolumn{1}{c}{$\chi^2/\nu$} &
\multicolumn{1}{c}{$\log_{10}{(X/{\rm Gyr^{-1}})}$} & 
\multicolumn{1}{c}{$\alpha$} &
\multicolumn{1}{c}{$\chi^2/\nu$} &
\multicolumn{1}{c}{$\log_{10}{(X/{\rm Gyr^{-1}})}$} & 
\multicolumn{1}{c}{$\alpha$} &
\multicolumn{1}{c}{$\chi^2/\nu$} 
\\
\hline
10.0	-	10.3	&	-1.09	$\pm$	0.24	&	-0.6	$\pm$	3.1	&		&	-0.82	$\pm$	0.19	&	-2.9	$\pm$	2.2	&	 	&	-1.59	$\pm$	0.29	&	1.9	$\pm$	3.9	&		\\
10.3	-	10.5	&	-1.38	$\pm$	0.13	&	2.9	$\pm$	1.2	&	0.59	&	-1.31	$\pm$	0.10	&	3.4	$\pm$	0.8	&	1.34	&	-1.85	$\pm$	0.34	&	2.2	$\pm$	4.7	&		\\
10.5	-	10.8	&	-1.57	$\pm$	0.09	&	4.6	$\pm$	0.6	&	0.34	&	-1.27	$\pm$	0.09	&	3.0	$\pm$	0.6	&	1.03	&	-2.32	$\pm$	0.11	&	8.5	$\pm$	0.9	&	4.35	\\
10.8	-	11.0	&	-1.64	$\pm$	0.09	&	4.2	$\pm$	0.6	&	0.33	&	-1.31	$\pm$	0.08	&	3.1	$\pm$	0.4	&	0.47	&	-2.42	$\pm$	0.14	&	6.6	$\pm$	0.9	&	0.58	\\
11.0	-	11.4	&	-1.74	$\pm$	0.11	&	4.1	$\pm$	0.6	&	0.34	&	-1.38	$\pm$	0.08	&	3.5	$\pm$	0.4	&	0.26	&	-2.32	$\pm$	0.12	&	4.9	$\pm$	0.6	&	1.22	\\
11.4	-	12.0	&	-1.95	$\pm$	0.10	&	4.7	$\pm$	0.4	&	1.15	&	-1.46	$\pm$	0.13	&	3.7	$\pm$	0.5	&	0.34	&	-2.13	$\pm$	0.18	&	2.7	$\pm$	0.8	&	1.18	\\
\hline
			&				&	4.4	$\pm$	0.3	&	& 			-1.36$\pm$0.41		&	3.4	$\pm$	0.3 &			&				&	5.7	$\pm$	2.5	&		\\       
\end{tabular}							        

\caption{Fits to specific star-formation rates as function of $z$ in stellar mass bins. 
Fit is to the function $sSFR=X(1+z)^\alpha$. Averages and standard deviations for $\alpha$ from data with  $M>10^{10.5}M_\odot$ are shown below the line.  Calculated for all galaxies and separately for ``blue" (templates Sab-Sdm and starburst, $3\le j_2\le11$)  and ``red" galaxies (both E templates, $j_2\le2$). Reduced $\chi^2$ are quoted if the number of points used in the fit is more than 2. $\alpha$ for SAM model varies between 2.44 and 2.67 over same mass range.}\label{tab:zfits}
\end{table*}


We do the same analysis for late/blue (Figure \ref{fig:ssfvsz_blue}) and early/red galaxies (Figure \ref{fig:ssfvsz_red}) as before.  It is striking that for the late-type galaxies the variation with stellar mass seen in Figure \ref{fig:ssfrvsz} is considerably reduced and that  there is negligible difference between galaxies with $M>10^{10.5}M_\odot$. Assuming, therefore, that they are measuring the same relation, we average the fits in Table \ref{tab:zfits} to estimate $\log_{10}({\rm  sSSFR})/{\rm Gyr}^{-1}=-1.36(\pm0.20)+3.36(\pm0.16)\log_{10}{(1+z)}$, where the error bars are errors in the mean calculated from the standard deviations given in Table \ref{tab:zfits}.

\section{Discussion}
Our stacking analysis shows that the specific FIR luminosity declines as a function of increasing stellar mass and evolves strongly with redshift.  Before we consider the implications of this it is worth drawing together some of the caveats.  Having corrected for the optical incompleteness the remaining selection effects are expected to be weak and should only effect the lowest luminosity or highest redshift cells.  The bias against very obscured objects would act to {\em reduce} either of the basic trends we found.  The possible bias to underestimate faint fluxes exposed in the simulations may bias us to slightly shallower slopes of sSSFR vs stellar mass.  Sampling variance is minimised by our large survey volume and is included in our error bars which take into account the variation from field to field.  Excluding galaxies with signs of AGN in the optical SED may bias us against star-formation correlated with AGN activity, or to higher AGN contamination in more massive systems where the signature of the AGN in the optical SED may be masked. The concordance between our independent analysis at 70 and 160 \micron\ also argues against any of these effects being significant.  Our biggest remaining caveat is that we are strongly reliant on the accuracy of the photometric redshifts and resulting stellar masses from \cite{RowanRobinson07} but we have mitigated against systematic errors in these by applying conservative constraints on the photometric redshifts and from our field-to-field comparison.

The specific FIR luminosity we measure includes energetic contributions from star formation, AGN and the ambient stellar radiation field which we relate to specific star formation rate.  Assuming these different contributions track each other with redshift then evolutionary trends will be unaffected.

With these caveats we consider that the trends we have observed in specific FIR luminosity mirror trends in specific star formation rates. The specific star formation rate is dependent on the factors triggering star formation, the availability of gas supplies and the importance of feedback processes that regulate the star formation.  The decline of specific star formation rate with stellar mass (which has already been observed at other wavelengths) may be due to declining resources, i.e. more massive galaxies have locked more of their baryons into stars in the past. Feedback also plays a role, e.g. AGN feedback in the more massive galaxies may suppress star formation (e.g. \citealt{bower,croton}). However, it seems that the abrupt changes in specific star-formation resulting from some models of AGN feedback (i.e. those of \citealt{Delucia2007}) are not supported. So either the impact of AGN feedback on specific star-formation is marginal or the AGN feedback is not as differentially dependent on stellar mass as in these models.

 The evolutionary trend is both remarkably consistent across different stellar masses in the range $10.5<\log_{10}M_*/M_\odot<12$, and stronger than the models. This discrepancy between the models and data could arise from an evolution in the feedback prescription which would need to be less restrictive at higher redshift.  Alternatively the triggering mechanisms (e.g. environmental effects) or decline in gas supplies in the models may need adjusting. It is worth noticing that enhanced evolution of the star formation rates enhances the change in specific star formation both through the star-formation directly but also through the more rapid stellar build-up.
  
It should also be noted that the parameters in the models including the ones controlling 
the AGN feedback strength, were only adjusted to reproduce galaxy properties at redshift zero and any resemblance at all to observed data at high redshift might be viewed as a partial success.

The presence or absence of a break in the specific star-formation rate as a function of stellar mass and the evolution of this with redshift appears to place significant constraints on feedback models. Our large sample means we have small statistical errors and so present results in finer redshift and mass bins. This means our results are less subject to changes across a bin. Our work extends previous work to lower redshift and has reduced errors at higher mass. By using the longer Spitzer wavelength and comparing 70 and 160\micron\ we provide a better constraint on the bolometric FIR power.  Our large area means we cover a representative range of environments (so are not subject to sampling variance problems) and are better able to assess systematic errors by comparing independent fields.   However, we have to compare with deeper samples to probe lower mass galaxies.  It is worth emphasising  that the stacking technique can overcome the problems of low sensitivity of far infrared data.  Given the availability of wide area far infrared and submm surveys from Spitzer and in the future from Herschel and SCUBA-2  a homogeneous analysis over representative samples of lower stellar mass would be possible with deeper optical data.

 \section{Conclusion}
 
 \begin{itemize}
\item We have used an improved stacking analysis to probe the far-infrared emission of galaxies near the peak of the cosmic infrared background.  

\item We have shown that SWIRE sources with $S_{3.6 \micron}$ contribute 70-80\% of the cosmic infrared background with the main uncertainty in the determination of background itself.

\item We show that about 50\% of the CIRB can be explored with these techniques to $r<23.5$, arguing that deeper optical data will have a dramatic impact on the science that can be done in these fields.

\item We have measured the average specific FIR luminosity or specific star formation rate as a function of stellar mass and redshift.

\item We have found a trend ${\rm sSFR} \propto M_* ^\beta$ with $\beta \sim -0.38$. This contrasts with SAMs in which the specific star-formation rate is constant with mass until a dramatic drop at high mass.  This trend is stronger for early type galaxies ($\beta \sim -0.46$) than late type galaxies ($\beta \sim -0.15$). 

\item We have found a strong evolutionary trend 
${\rm sSFR} \propto (1+z) ^\alpha$ 
with 
$\alpha = 4.4\pm0.3$ for 
$10.5<\log_{10}M_*/M_\odot\le 12$, steeper than a semi-analytic model which has $\alpha\sim2.4$.
.  
\item For early type galaxies the average evolution in this mass range is stronger ($\alpha\sim 5.7$) but decreases to higher mass. 
\item For late-type galaxies the trend is weaker but apparently independent of stellar mass, giving a mean rate $\alpha=3.36\pm0.16$.

\end{itemize}

\section*{Acknowledgments}
We thank the referee for useful comments, speedy responses and infinite patience. 

We thank Guilaine Lagache for useful discussion on stacking techniques.

Oliver, Farrah, Gonzalez-Solares, Babbedge and Roseboom have benefited from funding from the STFC (including grants PP/C502214/1, ST/F002858/1).
Frost acknowledges an STFC studentship.  
Henriques acknowledges his PhD fellowship from the Portuguese Science and
Technology foundation.

We have made use of the Ned Wright's Cosmological Calculator \cite{2006PASP..118.1711W}

This work is based on observations made with
the Spitzer Space Telescope,
 which is operated by the Jet Propulsion Laboratory,
California Institute of Technology under a contract with NASA.

Optical data used for photometric redshifts has come from the Isaac Newton Telescope and Palomar Observatory.

We have benefited from using TOPCAT \verb+http://www.star.bristol.ac.uk/~mbt/topcat/+

\appendix

\bsp

\label{lastpage}

\end{document}

%% file: table1.tex

\begin{tabular}{lrcccc}
\hline
Selection &  Number & Number Density & $I_{\rm 70}$ & $I_{\rm
  160}$ & $I_{\rm 70}/I_{\rm  160}$ \\
          &         & / 1000(sq. deg.)$^{-1}$ & /nWm$^{-2}$sr$^{-1}$ &
  /nWm$^{-2}$sr$^{-1}$ & \\
\hline
  $S_{36}>10\mu$Jy 	       & 394014	& 34.8	$\pm$1.6	& 4.231	$\pm$0.085	& 8.77	$\pm$0.44	& 0.48	$\pm$0.03 \\
$S_{24}>400\mu$Jy 	       & 21146	& 1.87	$\pm$0.05	& 2.159	$\pm$0.066	& 3.17	$\pm$0.06	& 0.68	$\pm$0.02 \\
$S_{70}>30$mJy		       & 856	& 0.076	$\pm$0.004	& 0.703	$\pm$0.039	& 0.68	$\pm$0.03	& 1.04	$\pm$0.08 \\
  $S_{160}>90$mJy	       & 877	& 0.077	$\pm$0.003	& 0.542	$\pm$0.041	& 0.71	$\pm$0.04	& 0.76	$\pm$0.07 \\
$S_{36}>10\mu$Jy; 	       		       		  	     	  		       			     	    
$13.5<r<23.5$ 		       & 223402	& 19.7	$\pm$0.6	& 3.247	$\pm$0.099	& 5.71	$\pm$0.32	& 0.57	$\pm$0.04 \\
$13.5\le r<23.5$; $\chi^2<5$;  		       		  	     	  		       			     	    
  $n_{\rm band}\ge4$ 	       & 182627	& 16.1	$\pm$1.0	& 2.390	$\pm$0.235	& 4.28	$\pm$0.38	& 0.56	$\pm$0.07 \\
  E ($j_2=1,2$) 	       & 29700	& 2.62	$\pm$0.09	& 0.260	$\pm$0.023	& 0.56	$\pm$0.05	& 0.47	$\pm$0.06 \\
  Sab ($j_2=3,4$) 	       & 9727	& 0.86	$\pm$0.05	& 0.198	$\pm$0.013	& 0.38	$\pm$0.02	& 0.52	$\pm$0.05 \\
Sbc ($j_2=5,6$) 	       & 23306	& 2.06	$\pm$0.12	& 0.417	$\pm$0.043	& 0.80	$\pm$0.06	& 0.52	$\pm$0.07 \\
Scd ($j_2=7,8$) 	       & 54864	& 4.84	$\pm$0.20	& 0.911	$\pm$0.070	& 1.63	$\pm$0.12	& 0.56	$\pm$0.06 \\
Sdm ($j_2=9,10$) 	       & 37594	& 3.32	$\pm$0.28	& 0.293	$\pm$0.056	& 0.45	$\pm$0.08	& 0.65	$\pm$0.17 \\
Star-burst ($j_2=11$) 	       & 23549	& 2.08	$\pm$0.17	& 0.276	$\pm$0.078	& 0.46	$\pm$0.08	& 0.60	$\pm$0.20 \\
AGN ($j_2=13-15$) 	       & 3887	& 0.34	$\pm$0.12	& 0.053	$\pm$0.019	& 0.06	$\pm$0.03	& 0.89	$\pm$0.51 \\
\hline
$S_{24}>60\mu$Jy               &        &                       & 5.94 $\pm$1.02        & 10.72 $\pm$ 2.28      \\
CIRB                           &        &                       & 6.4                   & 15.5  \\
\end{tabular}